\documentclass[acmsmall]{acmart}

\usepackage{graphicx}
\usepackage{comment}
\usepackage{color}
\usepackage{soul}
\usepackage{fancyvrb}
\usepackage{array}
\usepackage{tabularx}
\usepackage{booktabs}
\usepackage{cleveref}

\usepackage{tikz}
\usetikzlibrary{matrix}
\usetikzlibrary{shapes,arrows,positioning}
\usetikzlibrary{shadows.blur}
\usetikzlibrary{shapes.symbols}
\usetikzlibrary{shapes.geometric}
\usetikzlibrary{arrows,calc,positioning}
\usetikzlibrary{trees}
\usetikzlibrary{fit}

\tikzset{every shadow/.style={shadow opacity=40, shadow blur steps=7, shadow blur radius = .75pt, shadow xshift = 1.25pt, shadow yshift = -1.25pt}}
\tikzset{>=latex}

\definecolor{aaltoBlack}{RGB}{0,0,0}%
\definecolor{aaltoGray}{RGB}{146,139,129}%
\definecolor{aaltoRed}{RGB}{237,41,57}%
\definecolor{aaltoBlue}{RGB}{0,101,189}%
\definecolor{aaltoYellow}{RGB}{254,203,0}%
\definecolor{aaltoPurple}{RGB}{102,57,183}%
\definecolor{aaltoTurquoise}{RGB}{0,168,180}%
\definecolor{aaltoGreen}{RGB}{0,155,58}%
\definecolor{aaltoLightGreen}{RGB}{105,190,40}%
\definecolor{aaltoOrange}{RGB}{255,121,0}%
\definecolor{aaltoFuchsia}{RGB}{177,5,157}%

\usepackage{multirow}

\makeatletter
\def\zifour@scaled{s*[.95]}
\makeatother

\usepackage{subcaption}
\newcommand{\tablefontsize}{\fontsize{7.5pt}{9.5pt}\selectfont}

\usepackage{listings}
\newcommand\Smaller{\fontsize{6.75}{8.5}\selectfont}
\usepackage{xcolor}
\definecolor{aaltoBlack}{RGB}{0,0,0}%
\definecolor{aaltoGray}{RGB}{146,139,129}%
\definecolor{aaltoRed}{RGB}{237,41,57}%
\definecolor{aaltoBlue}{RGB}{0,101,189}%
\definecolor{aaltoYellow}{RGB}{254,203,0}%
\definecolor{aaltoPurple}{RGB}{102,57,183}%
\definecolor{aaltoTurquoise}{RGB}{0,168,180}%
\definecolor{aaltoGreen}{RGB}{0,155,58}%
\definecolor{aaltoLightGreen}{RGB}{105,190,40}%
\definecolor{aaltoOrange}{RGB}{255,121,0}%
\definecolor{aaltoFuchsia}{RGB}{177,5,157}%
\lstdefinestyle{cyaml}{
	aboveskip=0pt,
	belowskip=0pt,
	escapeinside={(*@}{@*)},
	keywordstyle=\color{aaltoBlue},
	keywords={containers,name,image,imagePullPolicy,command,args,containerPort,type,labels,port,protocol,app.kubernetes.io},
	classoffset=1,
	morekeywords={8080,apiVersion,kind,metadata,namespace,probe,selector,exclude,include,spec,ports},
	literate=%
    {0}{{{\color{aaltoOrange}0}}}1
    {1}{{{\color{aaltoOrange}1}}}1
    {2}{{{\color{aaltoOrange}2}}}1
    {3}{{{\color{aaltoOrange}3}}}1
    {4}{{{\color{aaltoOrange}4}}}1
    {5}{{{\color{aaltoOrange}5}}}1
    {6}{{{\color{aaltoOrange}6}}}1
    {7}{{{\color{aaltoOrange}7}}}1
    {8}{{{\color{aaltoOrange}8}}}1
    {9}{{{\color{aaltoOrange}9}}}1
}

\usepackage{enumitem}

\newcommand{\missed}{$\times$}%
\newcommand{\NA}{\textcolor{gray}{---}}%
\newcommand{\ppartial}{\LEFTcircle}
\newcommand{\found}{\CIRCLE}

\newcommand{\fakeparagraph}[1]{\vspace{.5em}\noindent\textbf{#1}.\hspace{0.25em}}
\renewcommand{\paragraph}[1]{\fakeparagraph{#1}}

\usepackage{wasysym}

\newcommand{\misconfiguration}[2]{\paragraph{#1: #2}}
\newcommand{\submisconfiguration}[2]{\textbf{#2 (#1)}}
\newcommand{\miscref}[1]{\textit{#1}}

\lstdefinestyle{KYAML}{
	aboveskip=0pt,
	belowskip=0pt,
	keywords={true,false,null,y,n,all,allow,Kubesonde},
	classoffset=1,
	keywordstyle=\color{aaltoBlue},
	morekeywords={apiVersion,kind,metadata,namespace,probe,selector,exclude,include,spec}
}

\def\totalapps{287}
\def\totalmisc{634}
\def\affectedapps{259}

\begin{document}

\title[Inside Job: Defending Kubernetes Clusters Against Network Misconfigurations]{Inside Job: Defending Kubernetes Clusters\\Against Network Misconfigurations}

\author{Jacopo Bufalino}
\affiliation{%
  \institution{CNAM}
  \city{Paris}
  \country{France}
}
\affiliation{%
  \institution{Aalto University}
  \city{Espoo}
  \country{Finland}
}
\email{jacopo.bufalino@lecnam.net}

\author{Jose Luis Martin-Navarro}
\affiliation{%
  \institution{Universitat Politècnica de València}
  \city{València}
  \country{Spain}
}
\affiliation{%
  \institution{Aalto University}
  \city{Espoo}
  \country{Finland}
}
\email{jose.martinnavarro@aalto.fi}

\author{Mario Di Francesco}
\affiliation{%
  \institution{Aalto University}
  \city{Espoo}
  \country{Finland}
}
\email{mario.di.francesco@aalto.fi}

\author{Tuomas Aura}
\affiliation{%
  \institution{Aalto University}
  \city{Espoo}
  \country{Finland}
}
\email{tuomas.aura@aalto.fi}



\begin{abstract}
Kubernetes has emerged as the de facto standard for container orchestration. Unfortunately, its increasing popularity has also made it an attractive target for malicious actors. Despite extensive research on securing Kubernetes, little attention has been paid to the impact of network configuration on the security of application deployments. This paper addresses this gap by conducting a comprehensive analysis of network misconfigurations in a Kubernetes cluster with specific reference to lateral movement. %
Accordingly, we carried out an extensive evaluation of \totalapps{} %
open-source applications belonging to six different organizations, ranging from IT companies and public entities to non-profits. As a result, we identified \totalmisc{} misconfigurations, well beyond what could be found by solutions in the state of the art. We responsibly disclosed our findings to the concerned organizations and engaged in a discussion to assess their severity. As of now, misconfigurations affecting more than thirty applications have been fixed with the mitigations we proposed.%
\end{abstract}

\maketitle

\section{Introduction}
\label{sec:introduction}

The security of cloud applications is a primary concern~\cite{Hussain:2020:techrep}. Protecting cloud environments from malicious actors presents unique challenges that are not only technology-dependent, but rather originate from the intrinsic characteristics of cloud-native applications -- those designed from the ground-up to take full advantage from the cloud paradigm~\cite{Rice:2020:book}. %
In fact, modern cloud-based applications comprise a large number of containerized microservices, that is, loosely coupled software components that leverage application programming interfaces (APIs) over a network~\cite{Newman:2021:book}. %
Kubernetes~\cite{Kubernetes:web} has become the de-facto standard for container orchestration in this context. 
Unfortunately, the security of Kubernetes applications is particularly worrying: a report recently released by Red Hat and based on a survey of 600 IT professionals revealed that 89\% of them experienced at least one security incident%
, even resulting in revenue or customer loss~\cite{RedHat:2024:web}.

\begin{figure*}[!tbp]
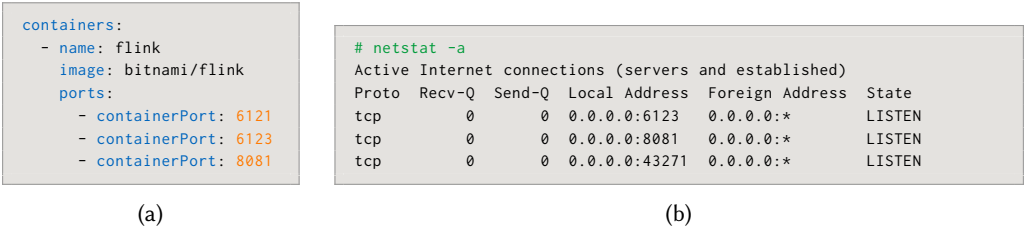

  \centering
  \subcaptionbox{\label{fig:misconfigurations:yaml}}[.2675\linewidth]{%
\begin{lstlisting}[style=cyaml]
containers:
  - name: flink
    image: bitnami/flink
    ports:
      - containerPort: 6121
      - containerPort: 6123
      - containerPort: 8081
\end{lstlisting}%
  }%
  \hspace{2em}%
  \subcaptionbox{\label{fig:misconfigurations:netstat}}[.6425\linewidth]{%
\begin{lstlisting}[language=bash]
# netstat -a
Active Internet connections (servers and established)
Proto  Recv-Q  Send-Q  Local Address  Foreign Address  State
tcp         0       0  0.0.0.0:6123   0.0.0.0:*        LISTEN
tcp         0       0  0.0.0.0:8081   0.0.0.0:*        LISTEN
tcp         0       0  0.0.0.0:43271  0.0.0.0:*        LISTEN
\end{lstlisting}%
  }%
  \caption{Configuration mismatch involving the Apache Flink application, a widely used data-flow framework for stream processing, as packaged by Bitnami. \subref{fig:misconfigurations:yaml} The corresponding Kubernetes configuration (i.e., YAML file) specifies 6,121, 6,123 and 8,081 as the container ports; \subref{fig:misconfigurations:netstat} the application is actually not listening to port 6,121 but uses an ephemeral port (i.e., 43,271), as reported by the \texttt{netstat} tool. These kinds of mismatches combined with application misbehavior may lead to different types of security issues %
  (see Sections~\ref{sec:examples} and~\ref{sec:misconfigurations:list}). %
  }\label{fig:misconfigurations}
\end{figure*}

Specifically, misconfigurations have been recognized as primary causes of attacks to Kubernetes~\cite{vmware:report:web,RedHat:2024:web,Dietrich:2018:SIGSAC,Acar:2017:SecDev}, also affecting Fortune 500 companies~\cite{aquasec:fortune500:web}. Misconfigurations occur when the deployment of Kubernetes application is poorly setup (Figure~\ref{fig:misconfigurations}), typically by administrators instead of the actual developers of these applications. %
The danger of such misconfigurations is amplified by the increasing trend of %
reusing third-party resources and configurations~\cite{cncf_helm_report}. Furthermore, many related issues arise from the cloud networking paradigm, which is substantially different from legacy corporate networks -- those using physical cables and on-premise hardware, including firewalls and switches, for segmentation purposes~\cite{Stallings:2015:book}. The Kubernetes model in particular is based on a flat network to simplify the transition of legacy applications to the cloud, however, its security implications are not fully understood~\cite{Minna:IEEE:2021}. As a consequence, it is challenging to %
correctly configure all the components of a %
cloud application, which remains a tedious and error-prone process%
~\cite{Rahman:2023:TSEM}.

Research on Kubernetes security has largely focused on hardening~\cite{Shamim:2020:SecDev}, risk assessment~\cite{Blaise:2022:CLOUD}, and tools for runtime analysis~\cite{MINNA2023}. Instead, studies on security misconfigurations have primarily targeted the cloud infrastructure~\cite{Rahman:2019:IST, Chiari:ICSA:2022,Banse:2021:CLOUD} and container runtimes~\cite{Jarkas:2025:ACS}. %
In fact, only a few works have explicitly addressed misconfigurations in the context of Kubernetes security~\cite{RedHat:2024:web,Rahman:2023:TSEM,Yang:CCS:2023,Gu:2025:SP}. The survey-based report in~\cite{RedHat:2024:web} established misconfigurations as one of the major factors behind security threats in Kubernetes, but did not further elaborate on their characteristics. The empirical study in~\cite{Rahman:2023:TSEM} targeted security misconfigurations in open-source Kubernetes applications, whereas the works of~\cite{Yang:CCS:2023,Gu:2025:SP} focused on excessive permissions granted to third-party components. %
However, the existing literature has paid little attention to security issues originated by network misconfigurations, in particular, those affecting actual cluster deployments (Section~\ref{sec:related-work}).

This work explicitly 
characterizes network misconfigurations affecting Kubernetes applications in the context of lateral movement (Section~\ref{sec:background}). %
Specifically, we rigorously examined research papers and industry artifacts as a basis for independent research that resulted in a comprehensive list of %
network misconfigurations (Section~\ref{sec:misconfigurations}). Then, we %
carried out an analysis of \totalapps{} open-source Kubernetes applications %
belonging to six different organizations, ranging from IT companies and public entities to non-profits. The analysis revealed a total of \totalmisc{} misconfigurations, most of which cannot be detected by solutions in the state of the art (Section~\ref{sec:evaluation}). Finally, we responsibly disclosed our findings to the respective organizations and assessed their severity. As of now, misconfigurations affecting more than thirty applications have been fixed with the mitigations we proposed (Section~\ref{sec:disclosure}).

In summary, the contributions of this paper are the following.
\begin{itemize}
    \item \textbf{Novel network misconfigurations leading to security issues}. Our analysis of Kubernetes cluster-internal networking unveiled different types of misconfigurations, well beyond what was known in the existing literature. We also show how these misconfigurations, together with application (mis)behavior, can lead to diverse security issues.
    \item \textbf{An evaluation of these misconfigurations in real-world application deployments}. We considered \totalapps{} open-source applications belonging to different organizations as deployed into a Kubernetes cluster. Our evaluation allowed to detect \totalmisc{} network misconfigurations. We also compare the effectiveness of our solution against the state of the art.
    \item \textbf{Disclosure and community engagement}. We disclosed the misconfigurations to the involved organizations and %
    explained how to mitigate them. We also carried out a security assessment and characterized the reliability of our results through the received feedback.
\end{itemize}

\section{Background}
\label{sec:background}

This section provides a few examples of attacks we discovered to motivate the study of network misconfigurations. It then gives an overview of Kubernetes and application deployment with Helm. %

\subsection{Motivating Examples: Security Issues Resulting from Network Misconfigurations}
\label{sec:examples}

We describe next two proof of concept attacks that are enabled by network misconfigurations.

\subsubsection{Broken Control Plane: Concourse}

Concourse~\cite{concourse:web} is a CI/CD software part of the CNCF landscape~\cite{landscapecncf}, consisting of a master web node and a number of different workers that are responsible for running builds and carrying out different types of resource checks. The Concourse application opens several dynamic ports in the ephemeral port range reserved by the host operating system (e.g., 32,768–60,999). These ports in the Concourse web node are endpoints of reverse SSH tunnels to worker nodes. The tunnels are command and control channels to the workers which should only be available at the loopback adapter of the web node. Instead, the ports are accessible from the cluster network due to the default Kubernetes behavior. Thus, any pod in the same cluster as the web node is able to send commands to a given worker, including other workers. As a result, we were able to deploy arbitrary containers, download container images, and edit running jobs.

\subsubsection{Service Impersonation: Thanos}

Thanos~\cite{thanos:web} is a set of components realizing high availability and long-term storage for the widely-used Prometheus monitoring system~\cite{Prometheus:web}. %
The related setup includes two compute units: \texttt{thanos-query-frontend}, responsible for external communication with users; and \texttt{thanos-query}, handling internal query processing. Both of them run different services that are associated with a single label, namely, \texttt{thanos-query-frontend}. As a consequence, an attacker inside either compute unit could exploit such a label to impersonate the service by simply listening to the appropriate ports. In fact, the load balancer or service targeting the resources bound to that label would send requests to the malicious pods in addition to (instead of) the legitimate ones, according to the applicable scheduling policy (i.e., depending on the specific proxy mode set in Kubernetes). Therefore, the attacker could fully impersonate the service, resulting in a denial of service (when the malicious pod does not respond) or even economic damage (by triggering unnecessary scale-up of cluster resources).

\subsection{Kubernetes and Container Networking}
\label{sec:kubernetes}

\begin{figure}[t!]
    \subcaptionbox{\label{fig:background:kubernetes}}{%
        \includegraphics[width=.525\textwidth]{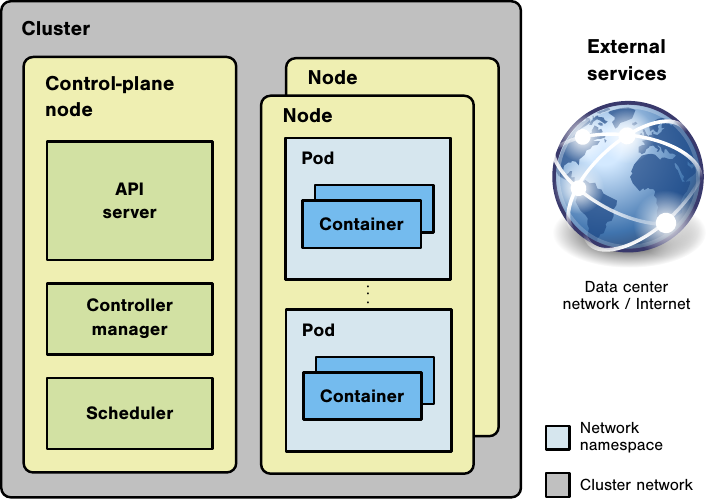}%
    }%
    \hfill%
    \subcaptionbox{\label{fig:background:helm}}[.4575\linewidth]{%
\begin{lstlisting}[style=cyaml,morekeywords={primary,service}]
# Helm manifest (values.yaml)
primary:
  service:
    type: ClusterIP
    ports:
      mysql: 3306
\end{lstlisting}\vspace{4pt}%
\begin{lstlisting}[style=cyaml]
# Helm template
kind: Service
metadata:
  labels:
    app.kubernetes.io/part-of: mysql
  ports:
    - name: mysql
      port: (*@\color{black}\{\{ .Values.primary.service.ports.mysql \}\}@*)
      protocol: TCP
\end{lstlisting}%
    }%
    \caption{\subref{fig:background:kubernetes} The main components in a cluster running microservices with Kubernetes. \subref{fig:background:helm} Sample fragments of a Helm manifest (top) and a Helm template (bottom). %
    The manifest defines values that can be referenced by the corresponding keys. The template in the example is a service and includes a directive (enclosed in double curly brackets) that sets the \texttt{port} key to the value of \texttt{mysql}. Such a value is specified as \texttt{.Values.primary.service.ports.mysql}, according to the structure of the manifest.
    }\label{fig:background}
\end{figure}

Kubernetes~\cite{Kubernetes:web} %
is the most widely used open-source software for container orchestration: it allows to deploy, scale and manage applications defined as interconnected microservices (Figure~\ref{fig:background:kubernetes}). Kubernetes relies on a set of \emph{nodes} (either bare-metal servers or virtual machines) together forming a \emph{cluster}. One node in the cluster has a special role: such a \emph{control-plane node} includes an API server (to interact with the orchestrator), a controller manager (to enforce desired attributes), and a scheduler (to allocate computing resources to nodes). The rest of the nodes, instead, run the actual applications (i.e., workloads). All resources in a cluster are specified by configuration files in the YAML format; computing resources are represented by software containers. Kubernetes objects are identified by their name and a list of key-value pairs called \emph{labels}~\cite{kubernetes:web:labels}. Kubernetes manages application lifecycle through a container \emph{runtime} (e.g., docker or podman) installed on the node to retrieve container images and create their execution environment. In particular, Kubernetes employs the abstraction of \emph{pod} to manage logically-related containers as an individual entity. 

Pods communicate with each other by exchanging messages over a network. The Kubernetes networking model is grounded on a few key abstractions~\cite{Minna:IEEE:2021}. Containers in the same pod are bridged to a shared Linux network namespace which allows communication over the same \emph{local network} (i.e., the so-called localhost). The address space in the cluster is flat, namely, pods can communicate with all other pods irrespective from the node where they are running. The \textit{service} resource acts as an abstract representation of a logical set of pods for load-balancing and service discovery, for instance. Pods are assigned to each service by means of their labels. Services can be cluster-internal or external, depending on the specific use case. Cluster-internal services are defined by the \texttt{ClusterIP} type and are assigned a unique IP address in the cluster network. As a special case, a \textit{headless} service~\cite{kubernetes:web:headless} allows to reach pods through the cluster's Domain Name System (DNS). %
Pods in the cluster can communicate with all services by default.

Kubernetes also supports %
network policies (i.e., through the \texttt{NetworkPolicy} object) to restrict access to pods and services. %
However, %
Kubernetes itself is not concerned about how actual IP addresses are managed and how network policies are enforced -- these are instead handled by external components called Container Network Interface (CNI) plugins. %

\subsection{Managing Kubernetes Applications} %
\label{ssec:helm}

In the context of Kubernetes, applications are just a collection of resources specified in YAML files, such as the sample in Figure~\ref{fig:misconfigurations:yaml}. %
Managing such files is tedious and error-prone, especially when the same resources need to be tailored for environments with varying characteristics, such as for development or testing purposes. %
For this reason, applications are usually specified through reusable Kubernetes configurations (generally called templates or blueprints) with specialized tools. Among them, Helm~\cite{helm:web} %
is a widely used software to package and manage applications for Kubernetes clusters; a Helm \emph{chart} is a collection of files that describes applications as Kubernetes resources, including configuration values and dependencies~\cite{helm:web:charts}. Specifically, a chart includes a manifest (specifying parameter values) and a template (describing a Kubernetes resource), as illustrated in Figure~\ref{fig:background:helm}. The templates are rendered into Kubernetes objects as YAML files by setting the parameters to the corresponding values. Helm charts are also composable, allowing to reuse other applications as dependencies -- many of them are publicly available and are specifically meant for sharing~\cite{artifacthub:web}. The terms (Kubernetes) application and (Helm) chart are used interchangeably in the rest of the paper.

\section{Network Misconfigurations}%
\label{sec:misconfigurations}

This section describes the network misconfigurations that can lead to security issues in Kubernetes. It first introduces the reference threat model, then explains the methodology employed to identify these misconfigurations. The section concludes with a detailed discussion of individual misconfigurations, along with an explanation of their causes and possible mitigations.

\subsection{Threat Model}
\label{sec:threat-model}

We explicitly refer to the threat matrix for Kubernetes by Microsoft~\cite{Microsoft:k8stm:web}, a widely-used resource in the context of container orchestration security, based on the MITRE ATT\&CK framework~\cite{mitre:web}. We restrict our attention to the lateral movement tactic in the threat matrix, particularly, the ``\emph{cluster-internal networking}'' technique therein. Specifically, we aim to identify and characterize misconfigurations possibly causing security threats in such a context for applications as they are deployed in a Kubernetes cluster. For this purpose, we assume that the attacker controls one container in a pod, which is aligned with previous work on Kubernetes security~\cite{Gu:2025:SP}. %
Moreover, our work assumes that %
the cluster has been hardened according to security best practices~\cite{clustersecurity:web}, namely, that it is not possible for the attacker to trivially take control of the whole cluster or a node. %
Finally, the container in control of the attacker has legitimate access to the cluster network but no other privileges that can be misused (e.g., root access or Kubernetes API access). %

\subsection{Methodology}
\label{sec:methodology}

We carried out a systematic review by collecting sources on network security for Kubernetes from academia and industry %
similar to the methodology in~\cite{Rahman:2023:TSEM,Ami:2022:SSP}, as detailed next. %
In particular, we determined misconfigurations based on the corresponding definition by the U.S.~National
Institute of Standards and Technology (NIST) in~\cite{Johnson:2011:NIST}, namely, ``\emph{an incorrect or suboptimal configuration of an information system or system component that may lead to vulnerabilities}''.

\subsubsection{Identifying Information Sources}

We collected research papers published in the past five years by searching both the Google Scholar and DBLP databases with the \emph{Kubernetes} and \emph{security} keywords; we specifically ensured full consideration of works published in top-tier conferences on both security (e.g., ACM CCS, IEEE S\&P, USENIX Security, and NDSS) and networking (e.g., CoNEXT, SIGCOMM, NSDI, and INFOCOM). Regarding industry sources, we went through security standards, whitepapers, and the Kubernetes documentation. In particular, we first considered all industry artifacts cited on the previously identified academic works: the CIS benchmark~\cite{cis-k8s-benchmark}, NSA CISA ``Kubernetes Hardening Guidance''~\cite{nsahardening2022}, OWASP~\cite{owaspmisconfig}, the Microsoft Threat Matrix for Kubernetes~\cite{Microsoft:k8stm:web}, the CNCF 2022 Annual Survey~\cite{cncfsurvey:web}. We then extended this corpus with public artifacts created in the last five years by Kubernetes-focused companies: VMWare ``State of Kubernetes''~\cite{vmware:report:web}, Red Hat ``2024 State of Kubernetes Security Report''~\cite{RedHat:2024:web} and ``Cryptojacking Attacks in Kubernetes''~\cite{RedHat:2020:web}, and Aquasec ``The State of Kubernetes and Docker Security in 2021''~\cite{aquasec:k8sreport:web}. We finally identified the software tools referenced in both academic and industry sources~\cite{Blaise:2022:CLOUD,RedHat:2024:web}.

\subsubsection{Inclusion and Exclusion Criteria}

We limited our attention to academic and industry artifacts specifically addressing network-related security issues or lateral movement in Kubernetes, and disregarded those that do not treat the topic in adequate detail or focus on other orchestrators, including related services offered by cloud providers~\cite{RedHat:2024:web,cis-k8s-benchmark,cncfsurvey:web,vmware:report:web,aquasec:k8sreport:web,Chiari:ICSA:2022}. As for software tools, we restricted our focus to those with each of the following properties:
they perform static or runtime analysis or are continuous monitoring tools (e.g., security platforms); they either support or specifically target Kubernetes; they are able to identify at least one type of issue related to networking. We specifically excluded: CNI plugins or tools that depend on a specific cloud provider, such as Tetragon~\cite{tetragon:web} and Azure Netpol manager~\cite{azurek8s:web}; policy engines or tools for compliance, such as Kyverno~\cite{kyverno:web} and Open Policy Agent~\cite{OPA:web}; software tools that are not publicly available, including AutoArmor~\cite{autoarmor} and Bastion~\cite{Nam:2020:ATC}; solutions specifically targeted for Infrastructure as Code, such as Tfsec~\cite{tfsec:web} and Tflint~\cite{tflint:web}.
As a result of this process, we obtained a list of research\,/\,industry artifacts~\cite{nsahardening2022,Microsoft:k8stm:web,owaspmisconfig,Kubesonde2023,Blaise:2022:CLOUD,Shamim:2020:SecDev,mahajan:IC3SIS:detection,MINNA2023,Kubescore:web,liu2022exploring,clustersecurity:web} and software tools~\cite{kubesec:web,checkov:web,kubeaudit:web,Kubescore:web,Rahman:2023:TSEM,kubebench:web,KubeScape:web,Trivy:web,stackrox:web,neuvector:web,KubeLinter:web}.

\subsubsection{Labeling Misconfigurations}

The resulting set of sources was leveraged to create a curated list of security issues broadly related to networking according to the principles of inductive thematic analysis~\cite{Braun:2021:book}. For this purpose, two of the authors independently labeled security issues into specific codes (namely, instances or types) of misconfigurations. Specifically, they started from an initial list of codes and compared them against the Kubernetes documentation, so as to identify gaps and opportunities to discover novel network misconfigurations. The latter entailed the creation of software tools as well as proof of concept attacks to selected Kubernetes applications (see Section~\ref{sec:examples} %
for more details). The outcome of this process was an initial set of labels for different misconfiguration types, which were then refined and consolidated in the list presented next.

\begin{table*}[!tbp]
  \caption{Identified network misconfigurations in Kubernetes, along with their security issues and possible attacks. Section~\ref{sec:misconfigurations:mitigation} discusses their mitigation.}\label{tab:misconfigurations}
  \def\arraystretch{1.125}
  \setlength{\tabcolsep}{4pt}
  \centering\sffamily%
  \tablefontsize\selectfont
  \centering
  \begin{tabular}{rlll}
    \toprule
    \textbf{ID} & \textbf{Description} & \textbf{Issue} & \textbf{Possible attack(s)} \\
    \cmidrule(lr){1-4}
    \multirow{2}{*}{\textbf{M1}}
      & \multirow{2}{*}{Port open on container is not declared}
      & \multirow{2}{*}{Listening on all interfaces by default}
      & Command and control \\
    & & & Sensitive port information \\
    \cmidrule(lr){1-4}
    \textbf{M2}
      & Container allocates dynamic ports
      & Dynamic ports cannot be controlled
      & Loosened security policies \\
    \cmidrule(lr){1-4}
    \multirow{2}{*}{\textbf{M3}}
      & \multirow{2}{*}{Port declared on container is not open}
      & \multirow{2}{*}{Missing checks on declared ports}
      & Data interception\,/\,spoofing \\
    & & & Data exfiltration \\
    \cmidrule(lr){1-4}
    \scalebox{.9}[1.0]{\textbf{M4A}}
      & Compute unit collision
      & \multirow{4}{*}{Missing checks on label collision} %
      & \multirow{4}{2.75cm}{Man in the middle\newline{}Server impersonation} \\
    \scalebox{.9}[1.0]{\textbf{M4B}}
      & Service label collision
      & 
      &  \\
    \scalebox{.9}[1.0]{\textbf{M4C}}
      & Compute unit subset collision
      & 
      & \\
    \textbf{M4*}\hspace{.425pt} 
      & Global label collision
      & 
      & \\
    \cmidrule(lr){1-4}
    \scalebox{.9}[1.0]{\textbf{M5A}}
      & Service targets unopened port
      & \multirow{4}{5cm}{Missing checks on declared ports\newline{}Missing checks on existence of target label} %
      & Data interception \\
    \scalebox{.9}[1.0]{\textbf{M5B}}
      & Service targets undeclared port
      & 
      & Data spoofing \\
    \scalebox{.9}[1.0]{\textbf{M5C}}
      & Headless service port is not available
      & 
      & Denial of service \\
    \scalebox{.9}[1.0]{\textbf{M5D}}
      & Service without target
      & 
      & Bypassing security checks \\
    \cmidrule(lr){1-4}
    \multirow{2}{*}{\textbf{M6}}
      & \multirow{2}{*}{Lack of network policies}
      & \multirow{2}{*}{No isolation between containers}
      & Data interception\,/\,spoofing \\
    & & & Privilege escalation \\
    \cmidrule(lr){1-4}
    \textbf{M7}
      & Container binds to host network
      & Network policies do not apply to host %
      & Bypassing network controls \\[2pt]
    \bottomrule
  \end{tabular}
\end{table*}

\subsection{Identified Misconfigurations}
\label{sec:misconfigurations:list}

This section details the identified misconfigurations (each associated with an identifier for convenience), including their possible impact on cluster security (Table~\ref{tab:misconfigurations}).  %
Those already recognized by the existing literature %
are referred to as \emph{Service without target} (\miscref{M5D}), \emph{Lack of network policies} (\miscref{M6}) and \emph{Container binds to host network} (\miscref{M7}); the rest were discovered through independent research.

\misconfiguration{M1}{Open ports are not declared}
It occurs when a container actually has ports open (at runtime) but these are not declared in the resource configuration (the Kubernetes YAML files). Open ports are %
available to all the pods by default, even when no access the host network is allowed, unless restrictive network policies are in place.
As a consequence, possible attacks include access to sensitive information as well as command and control. Representative applications exhibiting such security issues are those exposing administrative\,/\,debug ports and distributed services leveraging the worker-controller architectural pattern.

\misconfiguration{M2}{Dynamic ports}
Dynamic (also referred to as ephemeral) ports are communication endpoints whose identifier (i.e., number) is not explicitly defined by the developer. Instead, the operating system takes care of assigning a port number, which is chosen differently each time the associated server is started. Even though dynamic ports are not a misconfiguration per se, they cannot be specified as such in the declarative configuration of a Kubernetes application. For this reason, we treat them as a network misconfiguration. Even worse, dynamic ports cannot be completely blocked by enforcing network policies unless wide port ranges are specified. Sample applications using dynamic ports are those utilizing control\,/\,signaling channels (e.g., by multimedia protocols) or %
reverse TCP connections. %

\misconfiguration{M3}{Declared ports are not open}
It occurs when a certain port is specified in the declarative configuration of a container but the application is not actually listening to that port. Container ports can be arbitrarily opened and no check is performed on them by the orchestrator or at the lower layers. Combined with other weaknesses, this misconfiguration allows an attacker to carry out data spoofing, data interception, and data exfiltration just by listening to the declared port. Representative cases %
are applications supporting different deployment modes (for instance, running in a cluster or stand-alone) as well as scalable services that run in a single instance.

\misconfiguration{M4}{Label collisions}
They happen when unrelated resources are identified or targeted by the same set of labels. Collisions can be further classified based on the nature of the objects the labels are applied to. A \submisconfiguration{M4A}{Compute unit collision} takes place when the same label is applied to different compute units. Moreover, a \submisconfiguration{M4B}{Service label collision} occurs when multiple services target the same compute unit. Conversely, a \submisconfiguration{M4C}{Compute unit subset collision} involves unrelated compute units sharing some common labels which are selected as targets for a single service.
Collisions also affect different applications deployed into the same cluster; we refer to this case as a \submisconfiguration{M4*}{Global label collision}. Unfortunately, Kubernetes has no built-in functionality to prevent these: incorrect labeling possibly results in bypassing network policies and causes man-in-the-middle or server impersonation attacks. Sample applications exhibiting label collisions are those that use generic labels to describe resources.

\misconfiguration{M5}{Service with incorrect references}
It occurs when a service mistakenly refers to a given compute unit port; it does not generally pose security risks alone, however, it can amplify the reach of an attacker in the presence of other misconfigurations. There are different types of incorrect references, leading to a more detailed classification. One case is when a \submisconfiguration{M5A}{Service targets an unopened port}; it amplifies the spoofing or interception attack involving \miscref{M3} because services are the preferred way to contact applications (rather than compute units). This misconfiguration can also
cause a Denial of Service (DoS) as each request routed to that service will not receive a reply. Another type of misconfiguration manifests when a \submisconfiguration{M5B}{Service targets an undeclared port}. This might represent an evasion technique to circumvent static security checkers, for instance, to hide an SSH port. Moreover, there is the chance that a
\submisconfiguration{M5C}{Headless service port is not available}. A headless service is simply a DNS record that points to specific compute units in the cluster: a misconfiguration takes place when the service port differs from the one that opens in a container, resulting in mounting DoS (similar to \miscref{M5A}).
Finally, \submisconfiguration{M5D}{Services without target} occurs when a service does not match with any compute unit. This implies that every request involving such a service results in a failure. Similar to~\miscref{M4*}, a malicious user with access to the cluster could impersonate the service by deploying a compute unit with matching labels.

\misconfiguration{M6}{Lack of network policies}
We consider missing network policies as a misconfiguration, in alignment with the existing literature~\cite{cis-k8s-benchmark,Souppaya:2017:NIST}. 
Helm charts describing application deployments can specify network policies, but they can be enabled or not. In this case, we consider network policies that are available but not enabled as a misconfiguration. %

\misconfiguration{M7}{Container binds to the host network interface}
It occurs when the container bypasses the isolation provided by the Kubernetes networking layer and exposes ports to the underlying host network. This happens when the \texttt{hostNetwork} field in the declarative specification of a compute unit is set to \texttt{true}. As a consequence, the network namespace of the compute unit network becomes bound to the host network namespace, bypassing any network policy attached to it. %

\subsection{Causes and Consequences}
\label{sec:misconfigurations:causes-consequences}

The root cause of~\miscref{M1} and~\miscref{M3} is that the declarative nature of the ports is purely documentative, thus, it is not enforced by Kubernetes. This behavior aligns with the Kubernetes networking model; however, inaccurate characterization of ports is a major source of confusion for cluster administrators, especially for those reusing third-party charts. In fact, they would expect the configuration to be authoritative, to the point that they may use it to automatically setup network policies. Unfortunately, these would not properly work in presence of the two misconfigurations. Even worse, ports that are not accurately documented may affect the creation of network policies. Similar considerations apply to~\miscref{M5A} and~\miscref{M5B}. %
Somewhat related, the key issue behind \miscref{M2} is that dynamic ports are incompatible with the design principles behind Kubernetes networking~\cite{kubernetes:web:networking}. In fact, these ports cannot be practically specified in the configuration of a Kubernetes application.

The cause of~\miscref{M4} and all its variants is that Kubernetes lacks a mechanism to detect or prevent duplicate labels\,/\,selectors across compute units and services. %
Similarly,~\miscref{M5C} and~\miscref{M5D} occur because the orchestrator does not provide any warning that there is no target or port available for a given service. %
The main reason behind~\miscref{M6}, instead, is the default ``allow all'' connectivity policy in Kubernetes, which is clearly too permissive. In contrast, most Linux distributions and major cloud providers generally employ a ``deny all'' approach. %
Clearly, attackers can easily move laterally in the cluster if network access is not restricted.

Finally, the main motivation for~\miscref{M7} is the need to access hardware or operating system-related resources. This typically occurs when exposing metrics, as the Node Exporter component of Prometheus~\cite{Prometheous:node-exporter:web} does. Another example is given by scenarios that demand high efficiency, for instance, bandwidth-intensive applications such as those involving machine learning tasks~\cite{KubeDL:host-network:web}. Using the host network results in making network policies defined for a pod ineffective.

\subsection{Mitigation}%
\label{sec:misconfigurations:mitigation}

The last phase of our study entails deriving practical guidelines to avoid misconfigurations and therefore reduce their impact on possible security issues. %
We now discuss them in detail.

The mitigation of misconfigurations \miscref{M1}, \miscref{M3}, \miscref{M5A} and \miscref{M5B} is similar: it involves editing the configuration files to declare all the ports that are open in a container and to ensure that only such ports are bound to services. %
In particular, Helm charts demand special attention to handle ports that are open depending on specific parameter values (e.g., when optional features are enabled in the applications). As for \miscref{M5C}, it is enough to remove the port settings since headless services do not support such a feature. %
Handling misconfiguration \miscref{M2} is problematic, as it requires knowledge of the underlying application. The main objective is to declare the dynamic ports in some way. This can often be achieved by setting specific parameters in the applications to assign static ports %
instead of using the default (random) values. If this is not possible, the developer of the configuration should add a comment notifying their users that a certain application uses dynamic ports. This is needed because tools relying on network traffic analysis may erroneously generate policies for such ports, as they change every time the pod restarts. %
All \miscref{M4} variants can be addressed by making the labels unique. However, the process requires understanding the relationship between the different application components and the actual purpose behind using the same labels for them. %
Instead, \miscref{M5D} can be fixed by ensuring that each service has a selector matching the labels\footnote{This can be checked, for instance, with the \texttt{kubectl get pods -l <svc\_selector>} command.} of a compute unit. Mitigating \miscref{M6} requires defining\,/\,enabling network policies in the chart, making sure that each policy selects at least one pod and that rules only allow necessary connections. Finally, \miscref{M7} can be only remediated by setting \texttt{hostNetwork} to \texttt{false} after checking that such a change does not result in loss or functionality or performance. In the latter cases, alternative solutions should be sought or at least an in-depth security audit of the relevant pods should be carried out.

\section{Evaluation}
\label{sec:evaluation}

This section evaluates the proposed approach along different dimensions. First, it considers diverse datasets and carries out an analysis of the findings from running our solution. Second, it compares the effectiveness of the proposed approach against software tools in the state of the art.

\subsection{Sources} 
\label{sec:dataset}

We considered \totalapps{} open source Kubernetes applications defined as Helm charts by six different organizations. Accordingly, we divided them into different datasets as detailed next.

\subsubsection{Datasets}

We selected publicly available Kubernetes applications in the form of Helm Charts. We only picked applications from reputable sources and actively maintained. We also chose applications within specific organizations to represent diverse use cases, to gain a broad view of misconfigurations related to lateral movement. We classify and describe the datasets next. %
\begin{itemize}
  \item \emph{Sharing}. This category is given by charts defined by well-known organizations such as Bitnami (part of VMWare) and Banzai Cloud (owned by Cisco) which develop and officially support configurations meant to be reused and shared.
  \item \emph{Internal}. This class is represented by organizations maintaining Helm charts for their own software. %
  Specifically, we considered Wikimedia and the European Environmental Agency (EEA) as representative examples.
  \item \emph{Production}. The last group is represented by organizations such as the Cloud Native Computing Foundation (CNCF) and Prometheus Community which develop configurations for their own applications that are purposely built for production use. This use case is also representative of applications that are jointly deployed into the same Kubernetes cluster.
\end{itemize}

\subsubsection{Considered applications}

We selected only applications that run in their default configuration or with minimal changes for each dataset. Therefore, we did not consider applications that require specific environment variables, cloud providers, or images from private container registries. This is necessary to automate the testing process and to ensure the accuracy of the runtime analysis.

\subsection{Setup}
\label{sec:evaluation:setup}

We now briefly discuss the implementation of a solution that is able to identify the misconfigurations presented in the previous section. We also describe how we handled special cases of importance.

\subsubsection{Implementation}

We employed the characterization of the misconfigurations in Section~\ref{sec:misconfigurations:list} to create a list of machine-readable rules. We take a Helm chart as input then carry out both static and runtime analysis. We then combine the obtained results and evaluate them against the machine-readable rules. In doing so, we consider both the labels and the selectors associated with pods and services in addition to discrepancies between declaration and runtime behavior. Specifically, we carry out static analysis through a custom software that parses the YAML files and extracts the relevant information, including container ports, service ports, %
labels, and selectors. For the runtime analysis, we install each application into an empty Kubernetes cluster and observe its runtime behavior by following the approach in~\cite{Kubesonde2023}. In detail, we used Minikube~\cite{Minikube:web} version 1.23 and Kubernetes version 1.25. We opted for a virtualized environment to automate the process and ensure isolation. In particular, we delete and recreate the cluster after analyzing individual applications to guarantee that unrelated resources %
do not affect each other. Once all applications have been individually evaluated, we search for cluster-wide misconfigurations by checking the labels and selectors of every single application. %

\begin{table}[!tbp]
  \caption{Breakdown of network misconfigurations by dataset. In total, \totalmisc{} misconfigurations were found.}\label{tab:datasets_misc}
  \def\arraystretch{1.125}
  \setlength{\tabcolsep}{4.25pt}
  \centering\sffamily%
  \tablefontsize\selectfont
  \begin{tabular}{rrrrrrrrrrrrrrr}
    \toprule
      & 
      & 
      & 
      & 
      & \multicolumn{3}{c}{\textbf{M4}}
      & 
      & \multicolumn{4}{c}{\textbf{M5}}
      & 
      &  \\ \cmidrule(lr){6-8} \cmidrule(lr){10-13}
    \textbf{Dataset} & \scalebox{.95}[1.0]{\textbf{Affected apps}} & \textbf{M1} & \textbf{M2} & \textbf{M3}
      & \scalebox{.925}[1.0]{\textbf{M4A}}
      & \scalebox{.925}[1.0]{\textbf{M4B}}
      & \scalebox{.925}[1.0]{\textbf{M4C}}
      & \textbf{M4*}\hspace{-4pt}
      & \scalebox{.925}[1.0]{\textbf{M5A}}
      & \scalebox{.925}[1.0]{\textbf{M5B}}
      & \scalebox{.925}[1.0]{\textbf{M5C}}
      & \scalebox{.925}[1.0]{\textbf{M5D}}
      & \textbf{M6} & \textbf{M7}  \\
    \cmidrule(lr){1-1} \cmidrule(lr){2-2} \cmidrule(lr){3-15}
    Banzai Cloud   &  51\,/\,51   &  13 &   2 &  17 &   8 &   4 &   0 &   0 &   0 &   2 &   0 &   0 &  51 &   0 \\
    Bitnami        &  158\,/\,158 & 106 &  26 &  40 &  25 &  10 &   0 &   5 &   2 &  14 &   3 &   0 & 156 &   7 \\
    CNCF           &    7\,/\,10  &  10 &   0 &   4 &   0 &   0 &   0 &   0 &   6 &   0 &   0 &   0 &   7 &   0 \\
    EEA            &    8\,/\,19  &   7 &   0 &   1 &   0 &   1 &   0 &   0 &   0 &   0 &   0 &   0 &   0 &   0 \\
    Prometheus C.  &  25\,/\,25   & 42  &   4 &   3 &   0 &   0 &   0 &   0 &   1 &   4 &   0 &   0 &  25 &   4 \\
    Wikimedia      &   10\,/\,27  &  10 &   3 &   2 &   2 &   1 &   1 &   0 &   2 &   1 &   0 &   0 &   2 &   0 \\
    \cmidrule(lr){1-1} \cmidrule(lr){2-2} \cmidrule(lr){3-15}
    \emph{Total}         & 259\,/\,287 & 188 &  35 &  67 &  35 &  16 &   1 &   5 &  11 &  21 &   3 &   0 & 241 &  11 \\
    \bottomrule
  \end{tabular}
\end{table}

\subsubsection{Special Cases}

We also took additional steps to address a few corner cases that require special handling. %
The first one is related to the dynamic ports associated with~\miscref{M2}. These ports change every time the application is started, therefore, they are not captured by a single snapshot provided by the runtime analysis. To address such an issue, we perform the analysis twice then compare the ports detected in the two distinct iterations. We report \miscref{M2} when the detected ports differ. The second one occurs for \miscref{M7} because the application has access to the host network. In such a case, the runtime analysis reports all the ports that are open at the host, possibly including those of other components unrelated to the application. We address this issue by carrying out a preliminary analysis of the open ports in the host, which are then removed from the final output.

\subsection{Analysis}

The misconfigurations found by analyzing the considered applications %
are detailed next.

\subsubsection{Results}

\begin{figure}[!tbp]
  \subcaptionbox{\label{fig:misconfigurations:top10all}}{
    \includegraphics[height=3.3cm]{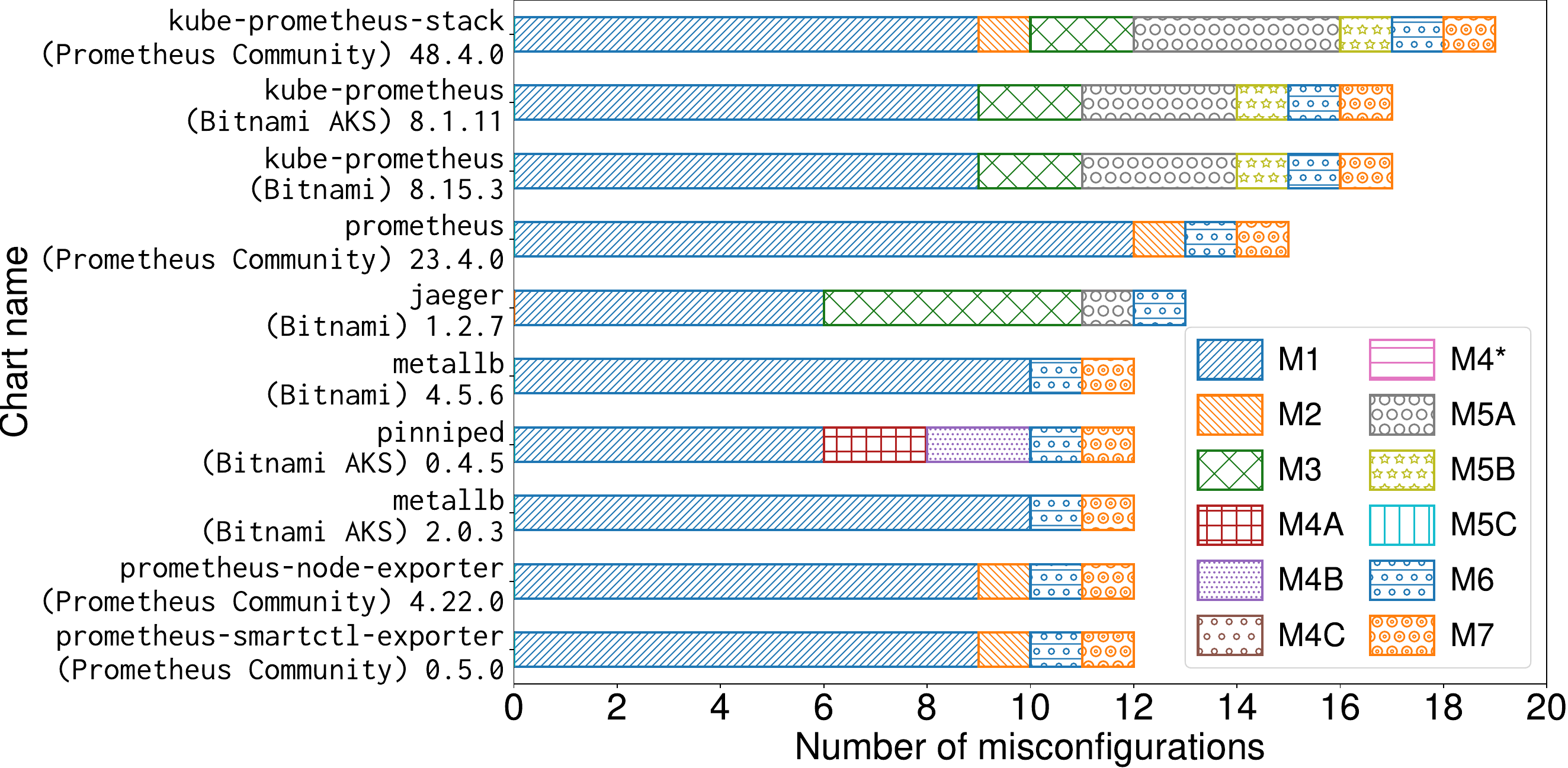}
  }%
  \subcaptionbox{\label{fig:misconfigurations:top10types}}{
    
  \includegraphics[height=3.2cm]{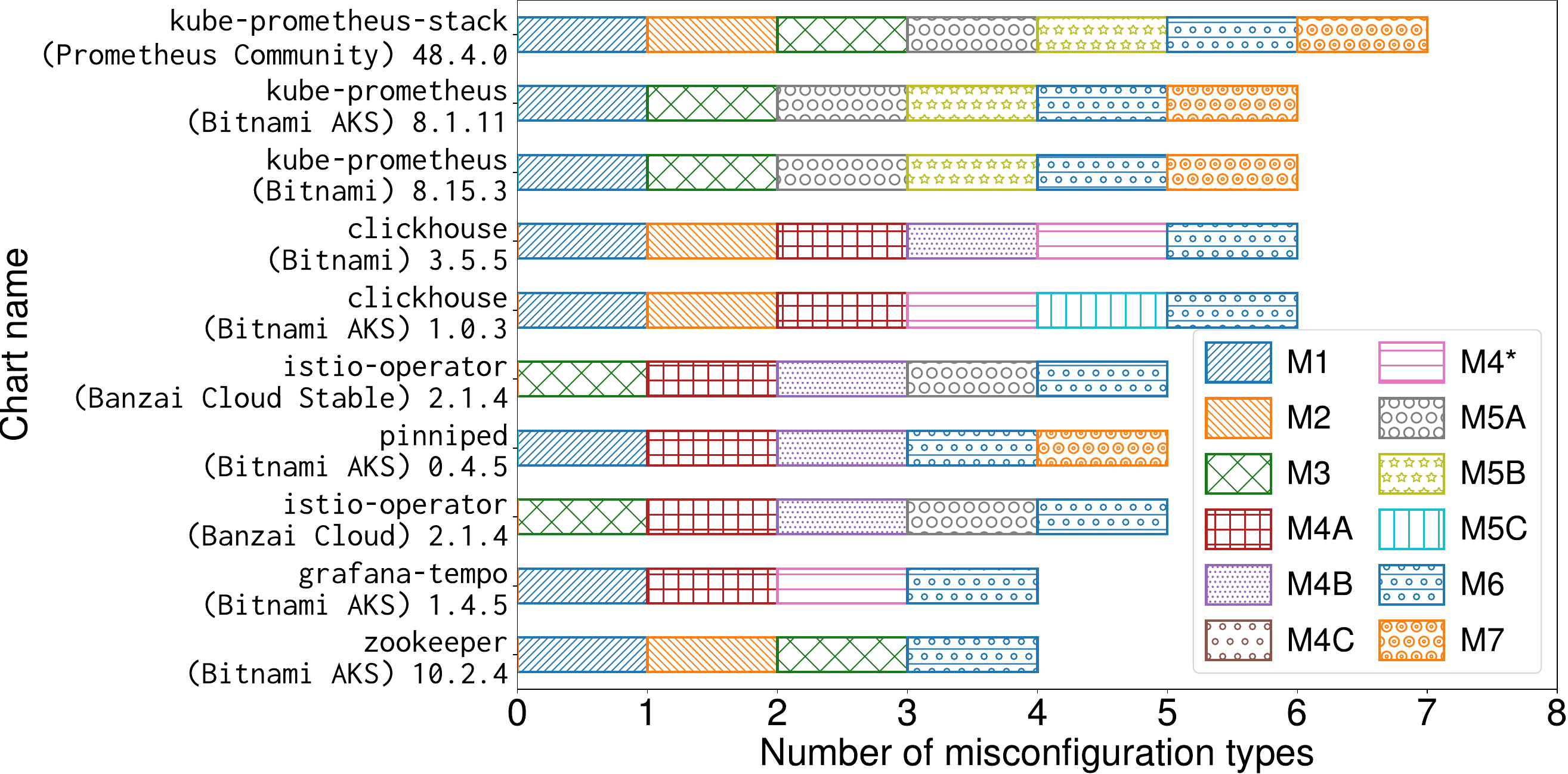}
  }%
  \caption{The ten applications with the highest number of \subref{fig:misconfigurations:top10all} misconfigurations and \subref{tab:results:network-policies} misconfiguration types.}\label{fig:misc}
\end{figure}

Table~\ref{tab:datasets_misc} details the outcome of our analysis, divided by dataset. As a result, 90\% (\affectedapps{} out of a total of \totalapps{}) of the applications in all the datasets have one or more misconfigurations. The most common %
misconfiguration types are \emph{Lack of network policies} (\miscref{M6}), \emph{Port not declared} (\miscref{M1}) and \emph{Port not open} (\miscref{M3}). These misconfigurations originate from lacking or inaccurate network boundaries on the Kubernetes resources, which could harm the cluster integrity. Approximately 10\% of the considered applications listen to ephemeral ports \emph{(M2)} and a similar number have label collision issues (\miscref{M4}). Among the label collision issues, the compute unit collision (\miscref{M4A}) is the most common, with 35 instances discovered. Service misconfigurations (\miscref{M5}) occur but less frequently; they only affect individual applications, not the cluster (those indicated as \miscref{M4*} in the table). Finally, only 3\% of applications have containers binding to the host network namespace (\miscref{M7}).  Figure~\ref{fig:misconfigurations:top10all} shows the ten most misconfigured applications and Figure ~\ref{fig:misconfigurations:top10types} shows the ten applications with the highest number of different concurrent misconfigurations.
These are all distributed applications spanning from log collection to load-balancing. Among the most misconfigured applications by number, all lack network policies (\miscref{M6}) and have multiple open ports that are not declared  (\miscref{M1}); 9 out of 10 bind to the host network namespace (\miscref{M7}); 4 out of 10 applications listen to ephemeral ports  (\miscref{M2}). Regarding the applications with the most types of misconfigurations, they also lack network policies but the other misconfigurations are more evenly distributed.

We then focused our attention on misconfigurations based on the type of dataset they belong to. For this purpose, we grouped the datasets according to their use case as described in %
Section~\ref{sec:dataset}. We noted that every application in the \emph{sharing} and \emph{production} datasets exhibited at least one misconfiguration -- their average number of misconfigurations per application was 3.35 and 4.44, respectively. In contrast, misconfigurations were found in only 39\% of charts built for \emph{internal} use, with an average number of misconfigurations slightly above one (specifically, 1.11). These results show that charts built by third parties have a significantly higher number of misconfigurations than those built for internal use: three to four times on average. This happens as charts are shared to promote adoption; as a consequence, their creators favor support for diverse functionality (e.g., optional application components) over strict enforcement of secure defaults. In contrast, charts intended for internal use are tightly integrated, whereas those for production environments generally rely on additional components or tooling. See also Section~\ref{ssec:feedback} for additional discussion. %

We lastly characterized how misconfigurations are distributed across the different applications in the datasets (Figure~\ref{fig:misconfigurations:histogram}). The data from our experiments revealed that 5\% of the applications exhibit at least 10 misconfigurations each, overall accounting for 25\% of the total number of misconfigurations. Moreover, 8\% of the applications had between 5 and 9 misconfigurations each, corresponding to about 22\% of the total. The rest of the misconfigurations were almost evenly distributed between the remaining applications. %
We have found that monitoring applications (such as observability frameworks) often exhibit open ports that are not declared (\miscref{M1}). Furthermore, applications that leverage a master-worker architecture (for instance, workflow engines) generally rely on dynamic ports (\emph{M2}) for coordination purposes. Finally, applications providing a substantial number of microservices (including those for federated services) tend to have declared ports that are not opened (\emph{M3}), because most of the corresponding features are not actually enabled at runtime.

\subsubsection{Impact of Network Policies}

\begin{figure}[!tbp]
  \subcaptionbox{\label{fig:misconfigurations:histogram}}{
    \includegraphics[height=3.5cm]{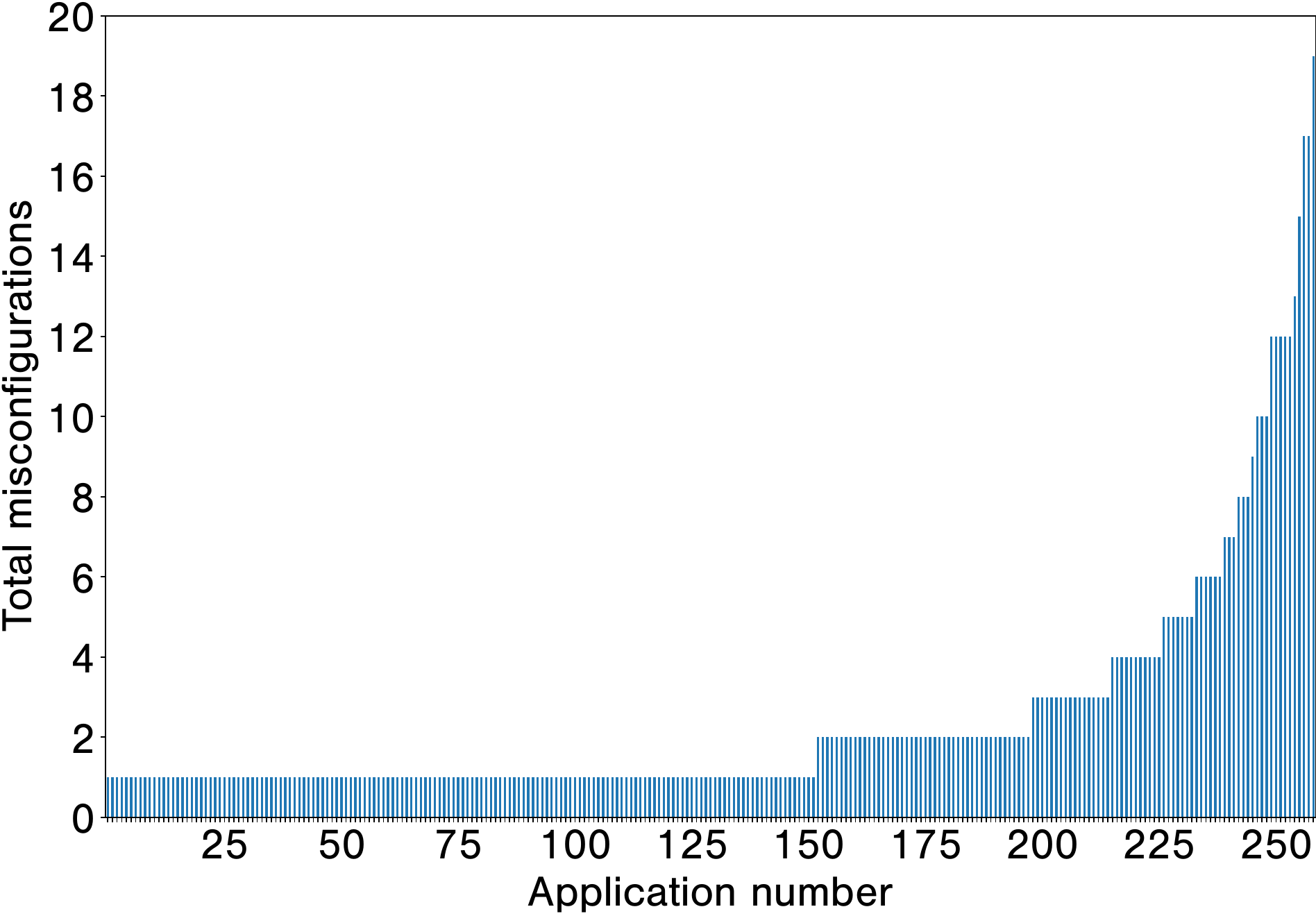}
  }%
  \hspace{.5cm}%
  \subcaptionbox{\label{tab:results:network-policies}}{%
    \bgroup
    \def\arraystretch{1.125}
    \centering\sffamily%
    \tablefontsize\selectfont
    \begin{tabular}{lrrrrrr}
      \toprule
      & \multicolumn{2}{c}{\textbf{Network Policies}} & \multicolumn{2}{c}{\textbf{Reachable}} \\
      \textbf{Dataset} & \textbf{Enabled} & \textbf{Affected} & \textbf{Pods} & \textbf{Services} \\
      \cmidrule(lr){1-1} \cmidrule(lr){2-3} \cmidrule(lr){4-5}
      Bitnami           & 48 (2) & 3 & 14 (1) & 0 \\
      CNCF              &  4 (0) & 0 &  0 (0) & 0 \\
      EEA               & 19 (19) & 8 & 13 (3) & 1 \\
      Prometheus C.     &  5 (0) & 3 & 32 (3) & 0 \\
      Wikimedia         & 25 (23) & 4 &  8 (5) & 2 \\[1pt]
      \bottomrule
    \end{tabular}
    \egroup
  }%
  \caption{\subref{fig:misconfigurations:histogram} Total misconfigurations per application. \subref{tab:results:network-policies} Impact of network policies on endpoint reachability.%
  }\label{fig:analysis}
\end{figure}

Network policies %
can be ill-defined, leading to a false sense of security. To evaluate their impact on misconfigurations, we analyzed all charts that define network policies and enabled them if they were not active by default. We followed the same methodology described in Section~\ref{sec:evaluation:setup} to determine whether network misconfigurations were mitigated by the policies. Specifically, we parsed the results of the runtime analysis and searched for endpoints corresponding to misconfigured ports that remained reachable from within the cluster. Figure~\ref{tab:results:network-policies} shows the obtained results; the Banzai Cloud dataset is not reported as none of the applications therein had any network policies. %
Misconfigured charts with loose network policies are indicated as affected in the table. The values within parentheses indicate the use of dynamic ports. %

The table %
shows that enabling network policies did not remedy misconfigurations in most cases. The allowed connections to misconfigured pods are higher than those to misconfigured services, which is aligned with the previous findings and also due to the presence of dynamic ports. Clearly, the sheer amount of reachable pods and services depends on the size of the applications, in terms of the number of containers therein. Also note that the results are obtained by deploying a single application in the cluster; the opportunities for reaching misconfigured ports would increase for multiple applications deployed at once. In summary, the datasets contain unwanted or unknown connections even when network policies are applied. This is not always due to dynamic ports, but it is also caused by erroneous %
settings. For instance, we noticed cases where strict policies targeted pods with access to the host network; however, network policies are not effective in such a case.

\subsection{Comparison with State of the Art}
\label{sec:comparison}

This section evaluates our solution against state-of-the-art security tools for Kubernetes. %

\subsubsection{Considered Solutions}

We considered representative tools among those found as described in Section~\ref{sec:methodology}. For clarity, we divide them in the categories detailed below.

\begin{itemize}
\item\emph{Static analysis}.
They analyze Kubernetes manifests, YAML files, and other static configurations before they are deployed in the cluster, comparing them against a database of best practices~\cite{kubesec:web,checkov:web,Rahman:2023:TSEM} or security guidelines~\cite{kubeaudit:web,Kubescore:web}, such as the CIS benchmark for Kubernetes~\cite{cis-k8s-benchmark}. %
They also offer actionable insights to improve the security of the deployments.
\item\emph{Runtime analysis}.
Analyze Kubernetes resources deployed in a cluster via the Kubernetes API~\cite{kubebench:web} or by inspecting running applications~\cite{Kubesonde2023}, providing a comprehensive view of the system based on all available cluster resources. Some tools%
~\cite{KubeHunter:web} %
impersonate malicious actors to assess application security. %
\item\emph{Other approaches}
A few tools support both static and runtime analysis~\cite{KubeScape:web,Trivy:web}, making them \emph{hybrid} solutions. Another class of tools is represented by those performing continuous security monitoring%
, which are also commonly known as \emph{security platforms}%
~\cite{stackrox:web,neuvector:web}. %
\end{itemize}

\subsubsection{Methodology}

We selected representative Kubernetes configurations exhibiting the misconfigurations in Table~\ref{tab:misconfigurations} for the evaluation by ensuring that the considered charts are representative of all the misconfiguration types identified in the paper. We employed this approach for convenience, to reduce the input provided to the different tools and %
to simplify the analysis of the issues reported in the output. %
We applied the configuration in a running Kubernetes cluster with the same setup as Section~\ref{sec:dataset} for the tools belonging to the runtime, hybrid, and platform categories.

Note that specific tools are unable to identify certain misconfigurations as a limitation intrinsic to their nature. For instance, tools only performing static analysis cannot discover issues occurring at runtime; conversely, those purely carrying out analysis at runtime are unable to detect misconfiguration due to cluster-wide collisions. We %
indicate these as ``not applicable'' in the table.

\begin{table*}[!tbp]
  \caption{Network misconfigurations detected by the considered tools and our solution. The symbols %
  indicate whether these were \found{} found, \ppartial{} partially found (i.e., either find less misconfigurations or require multiple runs), \missed{} missed, or \NA{} not applicable (i.e., could not be found by intrinsic limitations in the type of tool).
  }\label{tab:comparison}
  \def\arraystretch{1.125}
  \setlength{\tabcolsep}{4pt}
  \centering\sffamily%
  \tablefontsize\selectfont
  \begin{tabular}{lrrccccccccccccc}
    \toprule
      & 
      & 
      & 
      & 
      &
      & \multicolumn{3}{c}{\textbf{M4}}
      &  
      & \multicolumn{4}{c}{\textbf{M5}}
      &  \\ \cmidrule(lr){7-9} \cmidrule(lr){11-14}
    \textbf{Tool} & \textbf{Version} & \textbf{Type} & \textbf{M1} & \textbf{M2} & \textbf{M3}
      & \scalebox{.9}[1.0]{\textbf{M4A}}
      & \scalebox{.9}[1.0]{\textbf{M4B}}
      & \scalebox{.9}[1.0]{\textbf{M4C}}
      & \textbf{M4*}\hspace{-4pt}
      & \scalebox{.9}[1.0]{\textbf{M5A}}
      & \scalebox{.9}[1.0]{\textbf{M5B}}
      & \scalebox{.9}[1.0]{\textbf{M5C}}
      & \scalebox{.9}[1.0]{\textbf{M5D}}
      & \textbf{M6} & \textbf{M7}  \\
    \cmidrule(lr){1-1} \cmidrule(lr){2-2} \cmidrule(lr){3-3} \cmidrule(lr){4-16}
    Checkov~\cite{checkov:web}    & 3.2.23    & Static         & \NA          & \NA          & \NA          & \missed                            & \missed                            & \missed                            & \NA                                & \NA                                & \missed                            & \missed                            & \missed & \found & \found \\
    Kubeaudit~\cite{kubeaudit:web}   & 0.22.1   & Static         & \NA          & \NA          & \NA          & \missed                            & \missed                            & \missed                            & \NA                                & \NA                                & \missed                            & \missed                            & \missed & \found & \found \\
    KubeLinter~\cite{KubeLinter:web}   & 0.6.8  & Static         & \NA          & \NA          & \NA          & \missed                            & \missed                            & \missed                            & \NA                                & \NA                                & \missed                            & \missed                            & \found & \missed & \found \\

    Kube-score~\cite{Kubescore:web}  & 1.18.0   & Static         & \NA          & \NA          & \NA          & \missed                            & \missed                            & \missed                            & \NA                                & \NA                                & \missed                            & \missed                            & \found & \found & \missed\\
    Kubesec~\cite{kubesec:web}   & 2.14.0     & Static         & \NA          & \NA          & \NA          & \missed                            & \missed                            & \missed                            & \NA                                & \NA                                & \missed                            & \missed                            & \missed & \missed & \found \\
    SLI-KUBE ~\cite{Rahman:2023:TSEM}  & N/A   & Static         & \NA          & \NA          & \NA          & \missed                            & \missed                            & \missed                            & \NA                                & \NA                                & \missed                            & \missed                            & \missed & \missed & \found\\
    Kube-bench~\cite{kubebench:web}  & 0.7.1  & Runtime        & \missed      & \missed      & \missed      & \missed                            & \missed                            & \missed                            & \NA                                & \missed                            & \missed                            & \missed                             &\missed & \missed & \found     \\
    Kubescape~\cite{KubeScape:web}   & 3.0.3   & Hybrid         & \missed      & \missed      & \missed      & \ppartial                           & \ppartial                           & \ppartial                           & \missed                                & \missed                            & \missed                            & \missed                            & \missed & \found & \found \\
    Trivy~\cite{Trivy:web}     & 0.49.1     & Hybrid         & \missed      & \missed      & \missed      & \missed                            & \missed                            & \missed                            & \missed                                & \missed                            & \missed                            & \missed                            & \missed & \missed & \found \\
    Neuvector~\cite{neuvector:web}    & 5.3.0               & Platform      & \missed      & \missed       & \missed & \missed                            & \missed                            & \missed                            & \missed                                & \missed                            & \missed                            & \missed                            & \missed & \missed & \found \\
    StackRox~\cite{stackrox:web}     & 3.74.9              & Platform     & \missed      & \missed      & \missed      & \missed                            & \missed                            & \missed                            & \missed                                & \missed                            & \missed                            & \missed                            & \missed & \missed & \found \\[1pt]
    \cmidrule(lr){1-1} \cmidrule(lr){2-2} \cmidrule(lr){3-3} \cmidrule(lr){4-16}
    \textbf{Our solution}       & \NA   & Hybrid        & \found       & \found       & \ppartial    & \found                             & \found                             & \found                             & \found                             & \found                             & \found                             &       
    \found                             & \found                             & \found                             & \found                                                          \\[1pt]
    \bottomrule

  \end{tabular}
\end{table*}

\subsubsection{Results}

Table~\ref{tab:comparison} shows the misconfigurations identified by the considered tools, %
grouped by their type for convenience. First of all, we can see that none of the existing tools are able to at least partially recognize all misconfigurations, even though they successfully detect one or more of them -- the lack of network policies (\miscref{M6}) and host network mapping (\miscref{M7}) are the most recognized. %

Clearly, tools only supporting static analysis are (by their inherent nature) unable to identify most of the misconfigurations because they only inspect the configuration files. Even so, they still miss many of the misconfigurations they could possibly detect: all of those involving label collision (\miscref{M4}) and most of what relates to services (\miscref{M5}) -- only KubeLinter~\cite{KubeLinter:web} and Kube-score~\cite{Kubescore:web} identify services without target (\miscref{M5D}). %
In contrast, security platforms and hybrid tools could in principle recognize most of the misconfigurations considered in this work, as they also analyze the applications at runtime. However, their effectiveness is poor in practice: only Kubescape~\cite{KubeScape:web} reports deployments in which common labels are used for resources, hinting at issues involving label collisions.
Security platforms do get system information from the Kubernetes APIs and monitor processes and network sockets; however, they do not make any effort in notifying the user about potentially misconfigured resources, effectively delaying the detection of lateral movement after it has already occurred. In contrast, our approach was the only one to identify all the misconfigurations.

We also checked the source code of the considered tools to obtain more insights why misconfigurations were missed. We found that most of them do not check the relationship between Kubernetes resources of different types, for instance, between a service and a pod. This explains why static analysis tools were not able to identify label collisions and service-related issues as we instead do. Moreover, tools performing runtime analysis only query the configuration of resources from the API Server but do not actually inspect the runtime environment of the containers (for instance, open ports). Finally, most security platforms do not check for network-related settings but allow to record the entire traffic and possibly generate network policies based on it. Unfortunately, the latter is not enough as it can only be done assuming that all recorded traffic is intended and legitimate.

\section{Disclosure}
\label{sec:disclosure}

Our approach allows to identify diverse misconfigurations, however, it does not assess their severity. This is because such an assessment requires expert knowledge in the domain of the application as deployed in the Kubernetes cluster. For this reason, we reported the results of our analysis to the relevant organizations and engaged with their developers to understand the actual security implications of the misconfigurations in specific cases.

\subsection{Process and Follow-up}

We first performed a responsible disclosure of our findings to the organizations associated with the datasets we analyzed (see Table~\ref{tab:datasets_misc}), reporting a total of \totalmisc{} misconfigurations. %
We then engaged in an active and constructive discussion with these organizations so as to assess their severity through expert knowledge in the domain of the application.
Appendix~\ref{app:disclosure:methodology} provides additional details on the methodology used for the disclosure. %

\subsubsection{Acceptance and Severity Assessment}
\label{sec:disclosure:acceptance}

All developers recognized the issues we reported as misconfigurations. Those using software tools for security analysis also acknowledged that the issues identified in our work had not been detected before. Moreover, the respondents reported label collisions (\miscref{M4}) as the most critical and ``declared port not open'' (\miscref{M3}) as the least critical in terms of security risks. A few developers revealed that security issues related to lateral movement were internally classified as minor. This substantiates our premise that cluster-internal networking is a security threat that is often overlooked.

Regarding the severity of the reported misconfigurations, a respondent from the European Environment Agency explicitly stated in their feedback that the changes were security mitigations,
whereas a developer from Bitnami mentioned ``\emph{improved security of the application}'' as a benefit in their pull requests. %
Finally, a respondent from Wikimedia reported the outcomes of a security review of the reported misconfigurations; their feedback excluded the chance of remote code execution, but did not rule out other attacks. %
Overall, two of the respondents addressed (at least partially) the identified misconfigurations in less than 72 hours after receiving the report -- at the time of writing, misconfigurations affecting 30 applications have been fixed across five different organizations. %
Appendix~\ref{app:disclosure:impact} provides more details on the specific changes that fixed these misconfigurations.

\subsubsection{Accuracy and Analysis of Special Cases}
\label{sec:disclosure:accuracy}

Engaging with the developers also allowed us to detect false positives in our implementation, as well as to identify distinctive configurations for certain applications. %
One of the developers pointed out issues on %
UDP ports displayed as part of the misconfigurations. After investigating these, we found out that %
random UDP ports were sometimes reported as part of the runtime analysis. %
This kind of false positives amount to about 8\% of the total misconfigurations initially identified; they are not included in the final results summarized in Table~\ref{tab:datasets_misc}. %
We also received feedback regarding the service label collision misconfiguration, explaining a case where the collision was deliberate. It involved two services targeting a pool of database pods, wherein a service load-balances the requests to the pool of pods while the other one (headless) always retrieves the same instance, intended for write operations. %

A special use case was found in the Wikimedia dataset. Our analysis of the dataset showed that nearly half of the applications expose ports that are not declared. A careful inspection of the repository and of the tooling used by Wikimedia revealed a rather unconventional use of the port declaration, which we confirmed with the developers. In fact, Wikimedia charts only declare the ports of the applications that are needed for communication with other components, whereas undeclared ports are only intended to be used within the component. This is enforced with a custom tool that automatically generates network policies from the ports declared in the configuration files. The misconfigurations we identified in this scenario cannot actually be considered false positives but %
show how users rely on the declarative configuration of applications, as discussed in Section~\ref{sec:misconfigurations:causes-consequences}.

\subsection{Feedback}
\label{ssec:feedback}

This section summarizes the findings of our analysis that emerged as part of the disclosure process.%

\subsubsection{Implicit Security Assumptions} %

The recipients of the disclosures reacted differently to the reported misconfigurations. Organizations that create charts for third parties stated that the port information included in the chart description should not be used to create network policies. They rely on the assumption that users leveraging their charts create their network policies by manually inspecting connections with real traffic, as one of the respondents stated: ``\emph{Security sensitive users should probably start with `allow nothing' and add allowances based on real world traffic policies}''. %
The respondents agreed that network policies are needed. %
Those who did not include network policies with their applications justified their choice because of their peculiar features: ``\emph{I agree users probably should write network policies for their applications [\ldots] I am not very convinced they should be in the helm chart, as they are so specific to each user}''. They also suggested using external tools (such as third-party network plugins) to define and enforce network policies.

It is also interesting to notice how organizations that consume charts created by others value accurate declaration of port information, compared to those that create charts for their own applications. This is because they use such information to generate network policies. We are convinced that omitting network policies as part of applications because of the above-mentioned concerns is not only unsafe but fundamentally incorrect. Kubernetes, indeed, treats most of the network resources as interfaces, so that users can choose how to implement them. Therefore, the \texttt{NetworkPolicy} resource is a suitable candidate to provide a generic policy description that can later be implemented by other plugins.

\subsubsection{Information Hiding}

We have also demonstrated that users actually leverage the ambiguous nature of the port description to fit their own needs. This happened for Wikimedia, where the port description in the pods is used to automatically generate network policies. This case is interesting because it follows the principle of ``information hiding'', where an interface is created between modules, exposing public interfaces and hiding those that are private. %

\section{Discussion} 
\label{sec:discussion}

This section summarizes our findings, offers recommendations on networking-related best practices for cloud-native applications running in Kubernetes, and examines the limitations of our work.

\subsection{Findings}
\label{sec:discussion:findings}

\paragraph{Ambiguous or Unspecific Guidelines}
Following security best practices may create a false sense of security for both organizations and individual developers. This is especially because these best practices are not evenly developed. For instance, the CIS Kubernetes benchmark~\cite{cis-k8s-benchmark} contains 30 recommendations on how to secure the API server and 13 about role-based access control, but only two of them refer to network policies. To make it worse, these guidelines are ambiguous or not specific. In fact, they only state ``ensure that the [container networking interface] supports Network Policies'' (5.3.1) and ``ensure that all Namespaces have network policies'' (5.3.2). Although the recommendations are indeed valid, they fail to provide enough detail on how to correctly configure network policies. Moreover, many key aspects of internal networking are left out, including how to ensure that all the resources are correctly isolated, and it is the duty of the administrator to manually check them. As previously highlighted, Kubernetes does not automatically alert if a resource to be deployed has a label collision with an existing one in the cluster, or if the ports specified in a network policy actually match those of a given pod or service.

\paragraph{Reliance on Security Tools}
Organizations rely on security tools to assess the security of their Kubernetes clusters~\cite{RedHat:2024:web, vmware:report:web} and to alleviate the manual work required to secure the cluster. However, these tools often automate the checks suggested by the guidelines, inheriting their limitations and ambiguities. Therefore, they are able to correctly find inconsistencies on access control policies but miss most of the networking misconfigurations presented in this paper. %
Unfortunately, security tools are still immature~\cite{MINNA2023} and do not completely cover all the spectrum of misconfigurations that can appear in a complex environment such as a Kubernetes cluster. Therefore, we recommend developers not to fully depend on those tools but to actively incorporate the measures proposed here to assess the security of their applications and clusters.

\paragraph{Zero Trust and Service Meshes}
Zero trust is an emerging approach to security that moves away from the traditional concept of static network perimeter~\cite{Rose:2020:NIST}. In fact, it assumes that no implicit trust is granted to resources purely based on the physical\,/\,network location and ownership of a device. As a result, %
all communication flows should be strongly authenticated and authorized~\cite{Rais:2024:book}. In practice, zero trust in %
Kubernetes is achieved by employing a service mesh such as Istio~\cite{istio:web} -- a software layer that allows to monitor and control the information flow between different microservices, typically through sidecar proxies that authenticate, authorize and encrypt communications~\cite{Li:2019:SOSE}. 
The misconfigurations presented in this work can bypass the security mechanisms of the service-mesh. In fact, using a service mesh does not imply %
that the microservices in the application can \textit{only} be accessed through the mesh, since Kubernetes allows unbounded pod-to-pod communication by default. Moreover, if applications in the mesh are misconfigured, the related security issues will inevitably concern the mesh itself as well. Therefore, network policies are still needed to protect direct access between components and enable defense in depth%
~\cite{istio:bestpractices}.

\subsection{Recommendations}

\paragraph{End users}
Users need to understand the network properties and settings of the applications they deploy to ensure proper network segregation. This involves carefully reviewing the network ports associated with services and pods, as well as their relationships through labels and selectors. In addition, users should verify that network policies are available, enabled, and properly configured. A recommended practice is to use runtime analysis tools~\cite{Kubesonde2023,neuvector:web} to monitor open ports in the cluster and critically assess their necessity -- especially for optional or non-essential components. 

A special case involves the use of dynamic ports. Network policies cannot handle them, therefore, users should block access between a pod with dynamic ports and any unrelated service. A better alternative is to change the configuration to prevent the use of dynamic ports. It is often sufficient to manually configure the ports by setting environment variables or editing the default configuration.

\paragraph{Chart creators}
Creators should include all necessary network-related information in their charts for users to effectively deploy them. As a first step, they should document all ports that are open in the pods and services of their applications. Moreover, it is advisable not only to define but also to enable network policies by default in the chart, thereby establishing a more secure baseline for Kubernetes applications. The best way to add network policies is to use the default \texttt{NetworkPolicy} resource because it does not rely on third-party plugins. Finally, creators should carefully define templates and default parameters in their charts to ensure secure and reliable deployments.

\subsection{Limitations}%

Several factors may affect the accuracy and validity of our results. Some of them are due to our software implementation; in particular, the runtime analysis erroneously reports UDP ports as open in some cases (see Section~\ref{sec:disclosure:accuracy}). Moreover, it might miss ports that are not open during the analysis, for instance, those triggered by incoming traffic or due to port knocking techniques.

Our approach assumes that all network-related configuration is self-contained in a Helm chart, particularly, that it is defined through core Kubernetes components. However, users may rely on external tools or third-party Kubernetes resources (e.g., network plugins), as highlighted in Section~\ref{sec:disclosure:accuracy}. Another limitation of our approach is the lack of a ground truth. As a result, it cannot provide a conclusive security assessment of an application without expert knowledge (see Section~\ref{sec:disclosure:acceptance}). Instead, our analysis reports misconfigurations similar to what linters do.

Finally, our evaluation methodology relies on automatically deploying applications defined as Helm charts. However, some applications require manual configuration or rely on cloud provider-specific components to be successfully deployed; we could not consider these in our evaluation. Despite covering several diverse datasets, the results we obtained may not generalize to other scenarios, for instance, those not involving open-source applications.

\section{Related work}
\label{sec:related-work}

Defensive security in the cloud spans multiple layers of the network infrastructure~\cite{Jang:IEEE:2015,Billawa:2022:ARES}. The rest of this section focus on the existing literature that is more closely related to our work.

\paragraph{Network isolation for containers}
Nam et al.~\cite{Nam:2020:ATC} carry out a security analysis of container networking and devise a high-performance network stack that enforces security through a communication sandbox. Similarly, Nakata et al.~\cite{Nakata:2021:UCC} propose a sandboxing mechanism to improve network isolation for containers %
with a low overhead. In contrast, this work allows to harden network security of Kubernetes clusters without relying on additional components. Zhu and Gehrmann~\cite{Zhu:2022:COMSNETS} design and implement a system to generate AppArmor profiles for Kubernetes applications, including network rules, by collecting measurements at runtime. Their system requires legitimate traffic to produce meaningful results, whereas our work allows to identify unnecessary or malicious connections in a cluster purely based on applications' endpoints. %
Open Policy Agent \cite{OPA:web} is a general-purpose engine for policy enforcement; it offers native support for Kubernetes as an admission controller that intercepts requests to the orchestrator API and applies policies. Service meshes~\cite{Li:2019:SOSE} such as Istio allow to easily realize access control, encryption, and end-to-end authentication in a Kubernetes cluster. However, these solutions rely on %
correct policies (as noted in Section~\ref{sec:discussion:findings}), whereas our solution enables finding misconfigurations that can be easily overlooked.

\paragraph{Attacks to Kubernetes clusters}
Minna et al.~\cite{Minna:IEEE:2021} point out that the network abstractions employed by Kubernetes may result in unexpected attacks when users approach cloud security with a mental model derived from physical networks. As a step further, this work defines different types of misconfigurations and identifies them in real applications belonging to different use cases. %
Yang et al.~\cite{Yang:CCS:2023} demonstrate lateral movement attacks caused by excessive permissions on Kubernetes service accounts. 
Gu et al.~\cite{Gu:2025:SP} devise a solution to automatically detect such permissions that can be exploited to harm a Kubernetes cluster. In contrast, we rather focus on network misconfigurations that originate from Kubernetes abstractions rather than on loosely defined role-based access control rules. %
Ben David and Bremler-Barr~\cite{Ronen:CLOSER:2021} show that Kubernetes clusters are prone to Economic Denial of Sustainability (EDoS) attacks -- those causing economic damage through unnecessary use of resources -- triggered by its auto-scaling mechanism. Chamberlain et al.~\cite{Chamberlain:2025:POMACS} devise a model based on Markov decision processes to characterize EDoS attacks to Kubernetes. Our work does not address such a specific attack but considers network misconfigurations that can lead to diverse security issues, depending on the specific behavior of a Kubernetes application and its deployment.

\paragraph{Security misconfigurations}
Several studies address system configuration in the context of infrastructure as code~\cite{Rahman:2019:IST}. Similarly, Dietrich et al.~\cite{Dietrich:2018:SIGSAC} analyze the causes behind the occurrence of misconfigurations by surveying system operators. In contrast, our work specifically targets Kubernetes and does not rely on specific vendors or cloud providers. %
Shamim et al.~\cite{Shamim:2020:SecDev} accurately review the best practices for securing Kubernetes clusters. They advocate to change default configurations and set audit\,/\,network control policies, however, they do not provide details on how to do that. Instead, our work offers a comprehensive account of network misconfigurations, including guidelines on how to fix them. %
A report by Red Hat on the state of Kubernetes security based on a large-scale survey%
~\cite{RedHat:2024:web} has found that 40\% of the respondents experienced security issues related to misconfigurations. Unfortunately, the report does not include a detailed analysis on their nature%
, as we do here instead.
Rahman et al.~\cite{Rahman:2023:TSEM} carry out an empirical study of security misconfigurations by manually inspecting a dataset of open-source Kubernetes manifests, then validate their findings through a custom static analysis tool. Instead, we apply both static and runtime analysis, consider multiple datasets, and specifically focus on misconfigurations related to Kubernetes networking.

\paragraph{Analysis of Kubernetes applications}
Minna and Massacci~\cite{MINNA2023} survey runtime analysis tools for Kubernetes developed by both the research community and the industry; as a result, they identify orchestration security and resilience to internal threats as open challenges. We specifically target the latter in the context of lateral movement within a Kubernetes cluster. %
Blaise and Rebecchi~\cite{Blaise:2022:CLOUD} devise a graph-theoretic methodology to assess the security of microservice-based applications defined in terms of Helm charts. %
However, their work derives connectivity purely based on what is declared in the configuration files; instead, our solution also leverages runtime analysis for a more reliable characterization of network misconfigurations. %
Kubesonde~\cite{Kubesonde2023} employs runtime analysis to derive the network connectivity of Kubernetes applications. Specifically, it just reports such a connectivity but does not assist users in identifying possible issues related to it as we do in this work. %
Finally, there is a large amount of software tools %
that help users find security issues related to Kubernetes~\cite{kubesec:web,checkov:web,kubeaudit:web,Kubescore:web,Rahman:2023:TSEM,kubebench:web,KubeScape:web,Trivy:web,stackrox:web,neuvector:web,KubeLinter:web}. Unfortunately, they do not explicitly point out specific network misconfigurations or fail to identify most of them, as demonstrated by Table~\ref{tab:comparison}.

\section{Conclusion}
\label{sec:conclusion}

This work investigated security issues originated by network misconfigurations affecting Kubernetes clusters, with special focus to lateral movement. A systematic review formed the basis for us to identify novel network misconfigurations that could result in security threats.
Accordingly, we %
analyzed a large amount of open-source Kubernetes applications from several organizations. Our evaluation revealed that application deployments exhibit a significant amount of these misconfigurations. The related findings were disclosed to the relevant organizations, most of which acknowledged and fixed the misconfigurations. Our work presents easily identifiable sources of security issues%
, thereby helping to secure Kubernetes deployments and prevent practical attacks. 

There are still interesting directions for future research. We believe that applications would benefit from being specified through modular charts with clearly defined required\,/\,optional dependencies. This would allow to more precisely specify the intended network connectivity between components, which could then be leveraged to possibly derive network policies in a automated way. Indeed, Kubernetes is configuration-centric, yet it is still difficult for end users to understand the actual connectivity between different microservices. We expect visualization and monitoring tools to become more effective in assisting users by explicitly supporting network-related metrics and providing proactive advice on %
the resulting security issues, for instance, through machine learning.

\begin{acks}
This work was partially supported by: the \grantsponsor{AKA}{Research Council of Finland}{https://www.aka.fi/en} under grants number \grantnum{AKA345964}{345964} and \grantnum{AKA357533}{357533}; the INCIBE-UPV's Chair of Cybersecurity funded by the European Union under the NextGenerationEU initiative through the Spanish government's Plan de Recuperación, Transformación y Resiliencia.
\end{acks}

\bibliographystyle{ACM-Reference-Format}
\bibliography{insidejob.bib}

\appendix
\newpage

\section{Disclosure}
\label{app:disclosure}

This section details the methodology employed for the responsible disclosure as well as its impact.

\subsection{Methodology}
\label{app:disclosure:methodology}

We first identified the preferred private disclosure mechanism for each organization; if not specified, we contacted %
the maintainer of the chart. The disclosure included: a list of identified misconfigurations and the affected charts; the considered threat model (in Section~\ref{sec:threat-model}) and a description of each type of misconfiguration, including suggestions on mitigations. %
We then engaged in an active and constructive discussion with these organizations so as to assess their severity through expert knowledge in the domain of the application. Specifically, we asked to internally circulate an anonymous questionnaire regarding our methodology and the severity of the misconfigurations. The questions (in Figure~\ref{fig:questionnaire}) concerned the role and experience of the respondent, considerations on security handling (including the use of security software), and feedback on the reported misconfigurations. %

\begin{figure}[b!]%
\sffamily\fontsize{7.625pt}{9.275pt}\selectfont%
\begin{enumerate}%
    \item \textbf{What is the size of your organization, if applicable? (number of employees)}
    [\emph{Options: 1-99; 100-999; 1,000-4,999; 5000 or more; Not applicable}]
    \item \textbf{What is your current role?}
          Please describe your job position (e.g., Software Developer, SRE, DevOps Engineer) [\emph{Text}]
    \item \textbf{How long have you been using Helm?} [\emph{Options: Less than a year; 1-2 years; More than 2 years]}
    \item \textbf{Do you follow any guidelines to secure Helm Charts? If so, what are the main steps?} [\emph{Text}]
    \item \textbf{Do you use any software tools or services to check the security of Helm Charts? If so, which? }[\emph{Text}]
    \item \textbf{Compared to Charts created by your organization, do you handle third-party Helm Charts differently? (e.g., sanity checks, security checks, testing)}
          Third-party Charts are Charts created outside your organization. [\emph{Text}]
    \item \textbf{What do you think of the following statements:}
          Lateral movement in Kubernetes means that an attacker gets control of other pods after getting a foothold into the cluster. Port information in a Helm chart defines the network ports used by the application's services. This includes specifying port names, port numbers, and target port numbers.
          \begin{itemize}
              \item Detecting lateral movement in a Kubernetes cluster is a critical issue  \emph{[Options on a 5-point Likert scale]}
              \item I trust the port information in Helm Charts
              \emph{[Options on a 5-point Likert scale]}
          \end{itemize}
    \item \textbf{Do you use network policies with your cloud applications?}[\emph{Options: Yes; No}]
    \item \textbf{Why do you use network policies? What are their advantages and disadvantages?}[\emph{Text, only if Yes was answered%
    }]
    \item \textbf{Why you don't use network policies? What are their disadvantages?}[\emph{Text, only if No was answered%
    }]
    \item \textbf{Evaluate your agreement or disagreement with the following statements using the scale provided. Choose the response that best reflects your opinion on the following misconfigurations:}
          Undeclared ports are ports used by the container running in a pod, but not declared by the Chart\,/\,Pod. Unused ports are ports declared by the Chart\,/\,Pod but not used by the container.Label collision happens when different Kubernetes components use the same set of labels.
          \begin{itemize}
              \item Undeclared ports are critical security risk 
              \emph{[Options on a 5-point Likert scale]}
              \item Unused ports are a critical security risk 
              \emph{[Options on a 5-point Likert scale]}
              \item Label collision is a critical security risk
              \emph{[Options on a 5-point Likert scale]}
          \end{itemize}
    \item \textbf{Why they are not a critical security risk?}[\emph{Text, only shown if one of the previous answers to the critical risk was negative%
    }]
    \item \textbf{Did you receive a security report about Helm misconfigurations, including Undeclared ports, Unused ports and\,/\,or Label collision?}
          Undeclared ports are ports used by the container running in a pod, but not declared by the Chart\,/\,Pod. Unused ports are ports declared by the Chart\,/\,Pod but not used by the container. Label collision happens when different Kubernetes components use the same set of labels. %
          [\emph{Options: Yes; No}]
    \item \textbf{Are there false positives in the reported misconfigurations?}
          False positives are components with unused\,/\,undeclared ports, or components with the exact same set of labels, but as a result of a design choice. [\emph{Text}]
    \item \textbf{Evaluate your agreement or disagreement with the following statements using the scale provided. Choose the response that best reflects your opinion regarding the mitigation and detection of misconfigurations:}\begin{itemize}
              \item The proposed mitigations are useful
              \emph{[Options on a 5-point Likert scale]}
              \item I will use a tool to detect the reported misconfigurations
              \emph{[Options on a 5-point Likert scale]}
          \end{itemize}
    \item \textbf{If the proposed mitigations were not useful, what would be a better option?} [\emph{Text}]
    \item \textbf{Does the report reflect the status of your project? Leave here your feedback about the report} [\emph{Text}]
    \item \textbf{Please leave here any other feedback you may consider useful for our research} [\emph{Text}]
\end{enumerate}%
\caption{Content of the feedback questionnaire used to follow-up to the disclosure.}\label{fig:questionnaire}%
\end{figure}

\subsection{Impact}
\label{app:disclosure:impact}

We finally explain the impact of disclosing the misconfigurations we have found to the pertinent organizations %
at the time of writing. In particular, we show that the reported misconfigurations have been taken into serious consideration and also fixed in many cases.

\begin{figure}[!hbp]
    \centering
    \framebox{\includegraphics[width=0.475\textwidth]{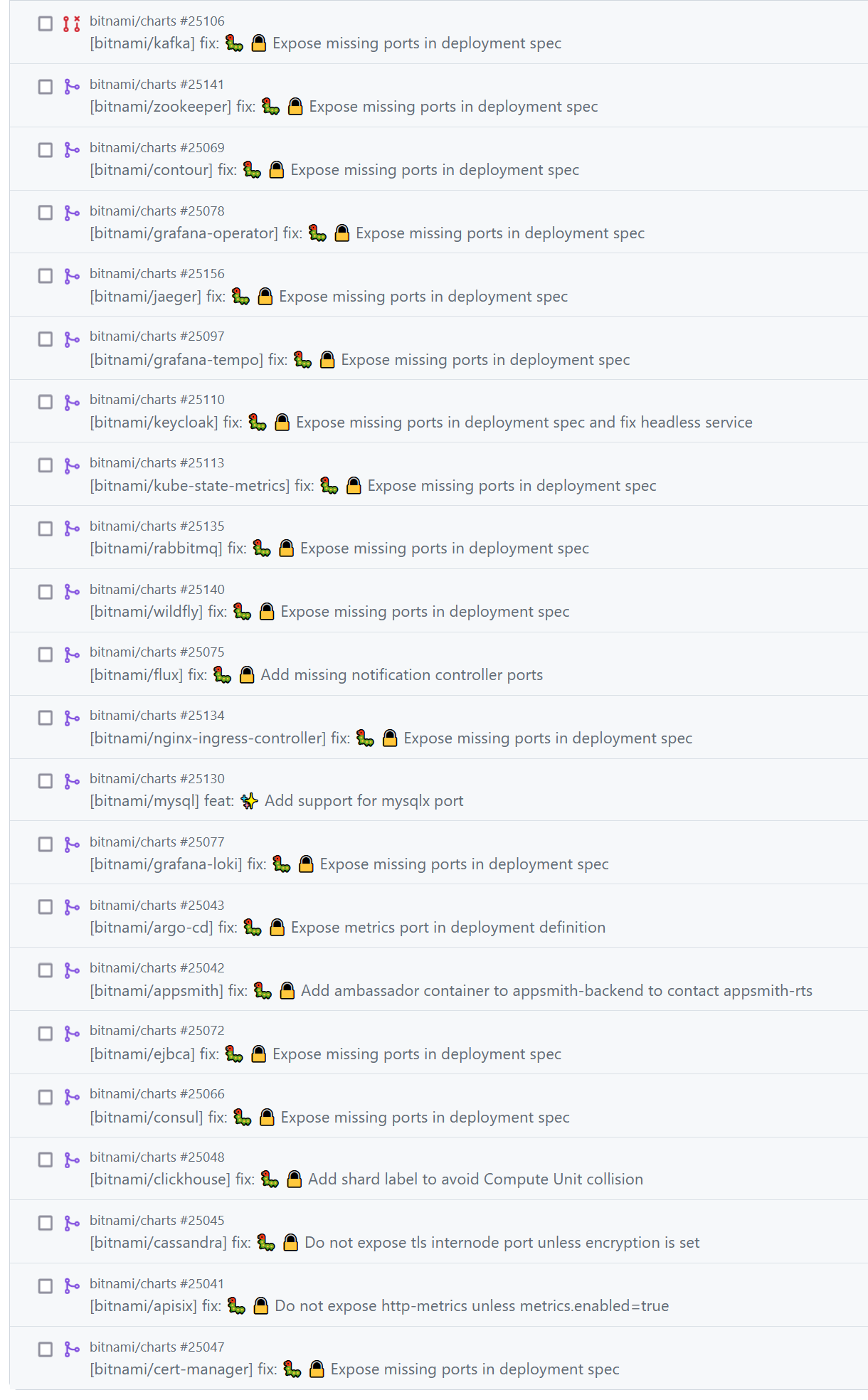}}
    \caption{Pull requests fixing misconfigurations in applications belonging to the Bitnami dataset.}\label{fig:pr-bitnami}
\end{figure}

\begin{figure*}[!tbp]
    \centering
    \framebox{\includegraphics[width=.8\textwidth]{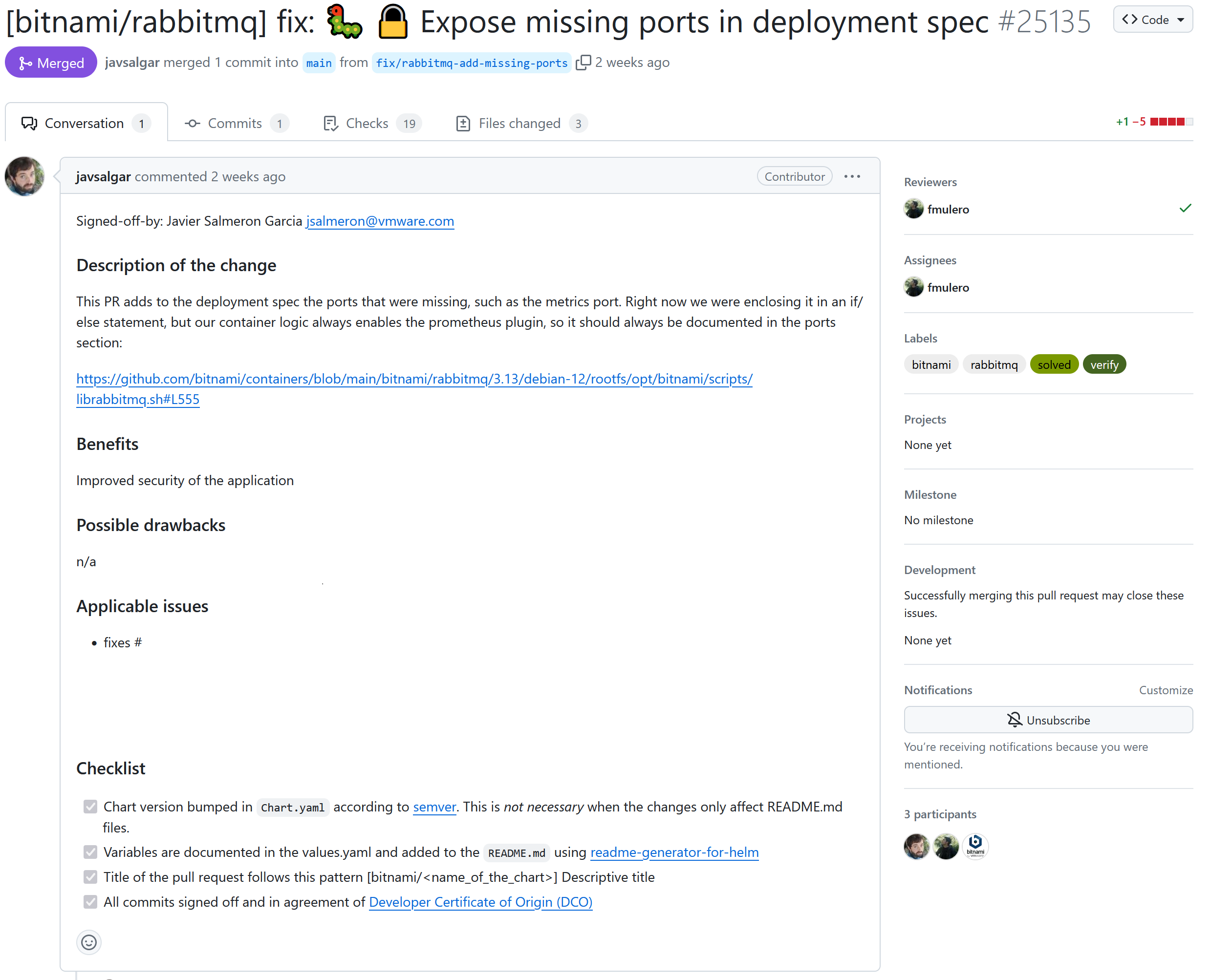}}
    \caption{Pull request fixing a %
    misconfiguration in the \texttt{rabbitmq} chart of the Bitnami dataset.}\label{fig:pr-bitnami-rabbitmq}
\end{figure*}

Figure~\ref{fig:pr-bitnami} illustrates that 22 pull requests have been created to fix the misconfigurations reported to Bitnami. In particular, Figure~\ref{fig:pr-bitnami-rabbitmq} shows a pull request for the RabbitMQ application -- a widely-used message broker -- to fix the \miscref{M1} misconfiguration (i.e., open ports are not declared).

\begin{figure}[!tbp]
    \centering
    \framebox{\includegraphics[width=0.475\textwidth]{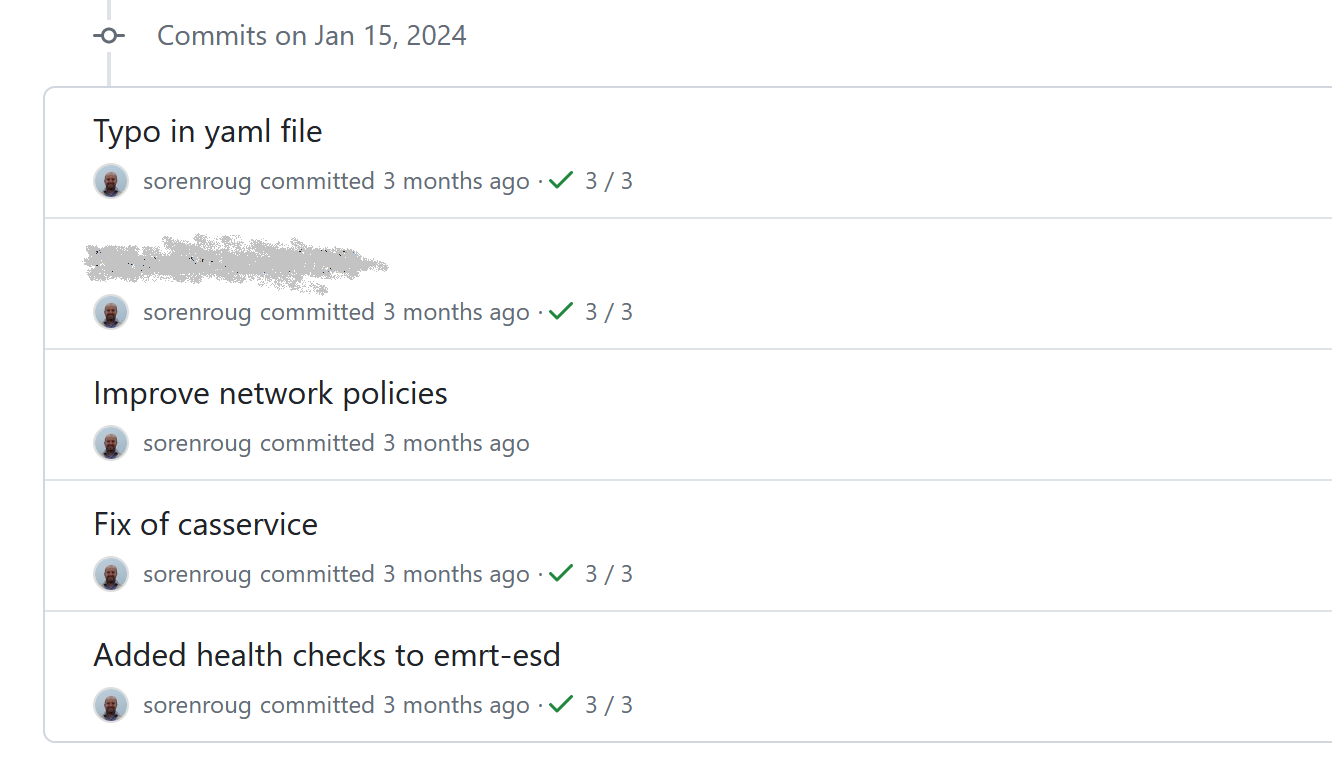}}
    \caption{Commits fixing misconfigurations in different charts belonging to the European Environment Agency (EEA) dataset.}\label{fig:pr-eea}
\end{figure}

\begin{figure*}[!tbp]
    \centering
    \framebox{\includegraphics[width=\textwidth]{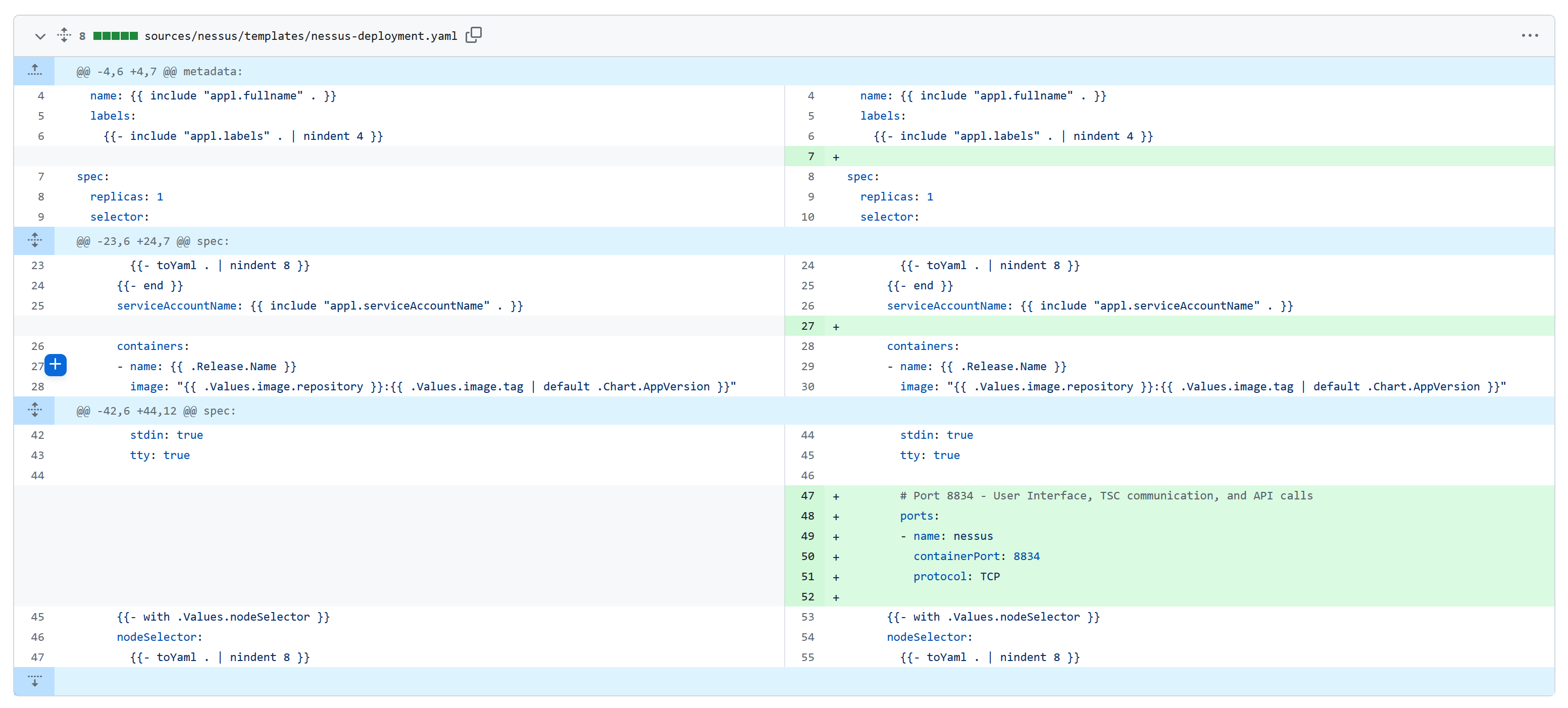}}
    \caption{Pull request fixing a %
    misconfiguration in the \texttt{nessus} chart of the EEA dataset.}\label{fig:pr-eea-fix}
\end{figure*}

Figure~\ref{fig:pr-eea} shows several commits fixing misconfigurations in different Helm Charts of the European Environment Agency (EEA). Figure~\ref{fig:pr-eea-fix} shows one of them, adding a missing port to the Helm chart.

\begin{figure*}[!tbp]
    \centering
    \framebox{\includegraphics[width=.8\textwidth]{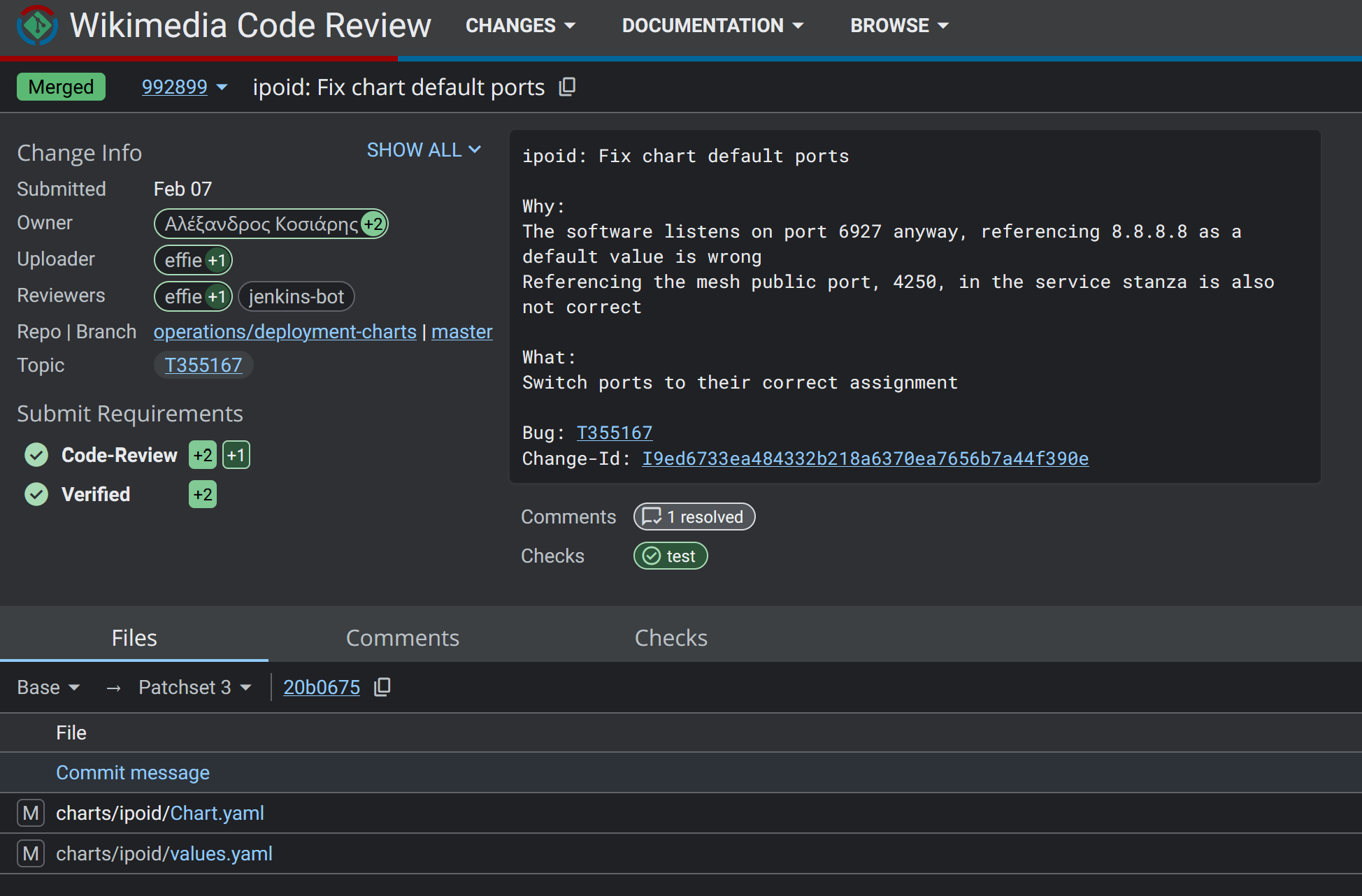}}
    \caption{Pull request fixing a misconfiguration in the \texttt{ipoid} chart of the Wikimedia foundation dataset.}\label{fig:wikimedia-pr}
\end{figure*}

The disclosure to the Wikimedia foundation resulted in four patches to fix some port declaration issues and the removal of unused functionality from other charts. Figure~\ref{fig:wikimedia-pr} is an example of the changes on one application, which listened on an incorrect IP address. %

\end{document}